\documentclass[12pt]{article}
\usepackage{latexsym}
\usepackage{times}
\usepackage{amsmath}
\usepackage{graphicx}
\usepackage{pict2e}
\usepackage{tikz}
\usepackage{pgfplots}
\usepackage{bussproofs}
\usepackage{booktabs}
\usepackage{adjustbox}
\usepackage{bm}
\pgfplotsset{compat=1.14}
\usepackage{url}
\usepackage{fullpage}
\newenvironment{acks}{}{}

\newcommand{\clabel}[1]{\makebox(0,0){#1}}

\newcommand{\llabel}[1]{\makebox(0,0)[l]{#1}}

\newcommand{\makenode}[1]{\bm{#1}}
\newcommand{\nodeu}{\makenode{u}}
\newcommand{\nodev}{\makenode{v}}
\newcommand{\nodew}{\makenode{w}}

\newcommand{\one}{\mbox{\bf 1}}
\newcommand{\zero}{\mbox{\bf 0}}
\newcommand{\leafone}{\makenode{T}_1}
\newcommand{\leafzero}{\makenode{T}_0}
\newcommand{\booland}{\land}
\newcommand{\boolor}{\lor}
\newcommand{\boolxor}{\oplus}

\newcommand{\tautology}{\top}
\newcommand{\nil}{\bot}
\newcommand{\obar}[1]{\overline{#1}}
\newcommand{\lit}{\ell}
\newcommand{\ite}{\mbox{\it ITE}}

\newcommand{\clausename}[1]{\textsc{#1}}
\newcommand{\cuhd}{\clausename{uhd}}

\newcommand{\culd}{\clausename{uld}}

\newcommand{\cvhd}{\clausename{vhd}}
\newcommand{\cvhu}{\clausename{vhu}}
\newcommand{\cvld}{\clausename{vld}}
\newcommand{\cvlu}{\clausename{vlu}}

\newcommand{\cwhu}{\clausename{whu}}

\newcommand{\cwlu}{\clausename{wlu}}
\newcommand{\candh}{\clausename{andh}}
\newcommand{\candl}{\clausename{andl}}
\newcommand{\cimh}{\clausename{imh}}
\newcommand{\ciml}{\clausename{iml}}

\newcommand{\opname}[1]{\mbox{\sc #1}}
\newcommand{\op}{\opname{Op}}
\newcommand{\andop}{\opname{And}}

\newcommand{\applyop}{\opname{Apply}}

\newcommand{\andchain}{A}

\newcommand{\cache}{\mbox{\it Cache}}

\newcommand{\turnstile}{\models}

\newcommand{\fname}[1]{\mbox{\small\sf #1}}

\newcommand{\lo}{\fname{Lo}}
\newcommand{\hi}{\fname{Hi}}
\newcommand{\var}{\fname{Var}}

\newcommand{\stephd}{\fname{HD}}
\newcommand{\stephu}{\fname{HU}}
\newcommand{\stepld}{\fname{LD}}
\newcommand{\steplu}{\fname{LU}}

\newcommand{\keyword}[1]{\textbf{#1}}
\newcommand{\keyif}{\keyword{if}}

\newcommand{\keyelse}{\keyword{else}}

\newcommand{\keyreturn}{\keyword{return}}
\newcommand{\assign}{\ensuremath{\longleftarrow}}

\definecolor{redorange}{rgb}{0.878431, 0.235294, 0.192157}
\definecolor{lightblue}{rgb}{0.552941, 0.72549, 0.792157}
\definecolor{clearyellow}{rgb}{0.964706, 0.745098, 0}
\definecolor{clearorange}{rgb}{0.917647, 0.462745, 0}
\definecolor{mildgray}{rgb}{0.54902, 0.509804, 0.47451}
\definecolor{softblue}{rgb}{0.643137, 0.858824, 0.909804}
\definecolor{bluegray}{rgb}{0.141176, 0.313725, 0.603922}
\definecolor{lightgreen}{rgb}{0.709804, 0.741176, 0}
\definecolor{redpurple}{rgb}{0.835294, 0, 0.196078}
\definecolor{midblue}{rgb}{0, 0.592157, 0.662745}
\definecolor{clearpurple}{rgb}{0.67451, 0.0784314, 0.352941}
\definecolor{browngreen}{rgb}{0.333333, 0.313725, 0.145098}
\definecolor{darkestpurple}{rgb}{0.396078, 0.113725, 0.196078}
\definecolor{greypurple}{rgb}{0.294118, 0.219608, 0.298039}
\definecolor{darkturquoise}{rgb}{0, 0.339216, 0.398039}
\definecolor{darkbrown}{rgb}{0.305882, 0.211765, 0.160784}
\definecolor{midgreen}{rgb}{0.560784, 0.6, 0.243137}
\definecolor{darkred}{rgb}{0.576471, 0.152941, 0.172549}
\definecolor{darkpurple}{rgb}{0.313725, 0.027451, 0.470588}
\definecolor{darkestblue}{rgb}{0, 0.156863, 0.333333}
\definecolor{lightpurple}{rgb}{0.776471, 0.690196, 0.737255}
\definecolor{softgreen}{rgb}{0.733333, 0.772549, 0.572549}
\definecolor{offwhite}{rgb}{0.839216, 0.823529, 0.768627}

\newcommand{\progname}[1]{{\sffamily\scshape #1}}
\newcommand{\kissat}{\progname{kissat}}
\newcommand{\Kissat}{\progname{Kissat}}
\newcommand{\buddy}{\progname{BuDDy}}

\newcommand{\tbuddy}{\progname{tbuddy}}
\newcommand{\Tbuddy}{\progname{Tbuddy}}
\newcommand{\tbsat}{\progname{tbsat}}
\newcommand{\Tbsat}{\progname{Tbsat}}
\newcommand{\drattrim}{\progname{drat-trim}}
\newcommand{\lratcheck}{\progname{lrat-check}}

\begin{document}

\title{Generating Extended Resolution Proofs \\ with a  BDD-Based SAT Solver \\ (Extended Version)}

\author{Randal E. Bryant \\ Marijn J. H. Heule \\
Carnegie Mellon University \\
Pittsburgh, PA 15213 USA \\
\texttt{\{Randy.Bryant,mheule\}@cs.cmu.edu}}



\begin{abstract}
In 2006, Biere, Jussila, and Sinz made the key observation that the
underlying logic behind algorithms for constructing Reduced, Ordered
Binary Decision Diagrams (BDDs) can be encoded as steps in a proof in
the {\em extended resolution} logical framework.  Through this, a
BDD-based Boolean satisfiability (SAT) solver can generate a checkable proof
of unsatisfiability.  Such a proof indicates that
the formula is truly unsatisfiable without requiring the user to trust
the BDD package or the SAT solver built on top of it.

We extend their work to enable arbitrary existential quantification of
the formula variables, a critical capability for BDD-based SAT
solvers.  We demonstrate the utility of this approach by applying a
BDD-based solver, implemented by extending an existing BDD package, to
several challenging Boolean satisfiability problems.  Our results
demonstrate scaling for parity formulas, as well as the Urquhart,
mutilated chessboard, and pigeonhole problems far beyond that of other
proof-generating SAT solvers.
\end{abstract}  




\maketitle

\section{Introduction}

When a Boolean satisfiability (SAT) solver returns a purported
solution to a Boolean formula, its validity can easily be checked by
making sure that the solution indeed satisfies the formula.  When the formula is
unsatisfiable, on the other hand, having the solver simply declare
this to be the case requires the user to have faith in the solver,
a complex piece of software that could well be flawed.  Indeed, modern solvers
employ a number of sophisticated
techniques to reduce the search space.  If one of those techniques is
invalid or incorrectly implemented, the solver may overlook
actual solutions and label a formula as unsatisfiable, even
when it is not.

With SAT solvers providing the foundation for a number of different
real-world tasks, this ``false negative'' outcome could have
unacceptable consequences.  For example, when used as part of a formal
verification system, the usual strategy is to encode some undesired
property of the system as a formula.
The SAT
solver is then used to determine whether some operation of the
system could lead to this undesirable property.  Having the solver
declare the formula to be unsatisfiable is an indication that the
undesirable behavior cannot occur, but only if the formula is truly
unsatisfiable.

Rather than requiring users to place their trust in a complex software
system, a {\em proof-generating} solver constructs a proof that the
formula is unsatisfiable.  The proof has a form that can
readily be checked by a simple proof checker. Initial work of checking unsatisfiability results
was based on resolution proofs, but modern checkers are based on stronger proof systems~\cite{ZhangMalik,appaHeule}.
The checker provides an independent validation that the formula is indeed unsatisfiable.  The
checker can even be simple enough to be formally verified~\cite{Lammich,cruz-cade-2017}.
Such a capability has become an essential feature for modern SAT solvers.

In their 2006 papers~\cite{Jussila:2006,ebddres}, Jussila, Sinz and Biere made the key observation that the
underlying logic behind algorithms for constructing Reduced, Ordered
Binary Decision Diagrams (BDDs)~\cite{Bryant:1986} can be encoded as steps in a proof
in the {\em extended resolution} (ER) logical framework~\cite{Tseitin:1983}.  Through this, a
BDD-based Boolean satisfiability solver can generate checkable
proofs of unsatisfiability.  Such proofs indicate
that the formula is truly unsatisfiable without requiring the user to
trust the BDD package or the SAT solver built on top of it.

In this paper, we refine these ideas to enable a full-featured,
BDD-based SAT solver.  Chief among these is the ability to perform
existential quantification on arbitrary variables.  (Jussila, Sinz,
and Biere~\cite{Jussila:2006} extended their original work~\cite{ebddres} to allow
existential quantification, but only for the root variable of a BDD.)
In addition, we allow greater flexibility in the choice of variable
ordering and the order in which conjunction and quantification
operations are performed.  This combination allows a wide range of
strategies for creating a sequence of BDD operations that, starting
with a set of input clauses, yield the BDD representation of the
constant function $\zero$, indicating that the formula is
unsatisfiable.  Using the extended-resolution proof framework, these
operations can generate a proof showing that the original set of
clauses logically implies the empty clause, providing a checkable
proof that the formula is unsatisfiable.

We evaluated the performance of both our SAT solver \tbsat{}
and \kissat{}, a state-of-the-art solver based on conflict detection
and clause learning (CDCL)~\cite{biere-kissat-2020,CDCL-handbook}.
Our results demonstrate that a
proof-generating BDD-based SAT solver has very different performance characteristics from the more mainstream CDCL solvers.  It does not do especially well as a general-purpose solver, but it
can achieve far better scaling for
several classic challenge problems~\cite{Alekhnovich,Chew,Haken:1985,Urquhart}.  
We
find that several of these problems can be efficiently solved using
the {\em bucket elimination} strategy~\cite{dechter-ai-1999} employed
by Jussila, Sinz, and Biere~\cite{Jussila:2006}, but others require a
novel approach inspired by symbolic model checking~\cite{burch-ic-1992}.

This paper assumes the reader has some background on BDDs and their algorithms.
This background can be obtained from a variety of tutorial presentations~\cite{andersen-1998,bryant-acm-1992,bryant-hmc-2018}. 
The paper is largely self-contained regarding proof generation.
The paper is structured as follows.  First, it provides a
brief
introduction to the resolution and extended resolution logical
frameworks and to BDDs.  Then we
show how a BDD-based SAT solver can generate proofs by augmenting algorithms for
computing the conjunction of two functions represented as BDDs, and
for checking that one function logically implies another.  We then
describe our  implementation and evaluate its performance on
several classic problems.  We conclude with some general observations and suggestions for further work.

This paper is an extended version of an earlier conference
paper~\cite{bryant:tacas:2021}.  Here we present more background material and more details about the proof-generation algorithms,
as well as updated benchmark results with a new implementation and on additional challenge problems.

\section{Preliminaries}

Given a Boolean formula over a set of variables $\{x_1, x_2, \ldots,
x_n\}$, a SAT solver attempts to find an assignment to these variables
that will satisfy the formula, or it declares the formula to be
unsatisfiable.  As is standard practice, A {\em literal} $\lit$ can be
either a variable or its complement.  Most SAT solvers use
Boolean formulas expressed in {\em conjunctive normal form}, where the
formula consists of a set of {\em clauses}, each consisting of a set
of literals.  Each clause is a disjunction: if an assignment sets any
of its literals to true, the clause is considered to be satisfied.
The overall formula is a conjunction: a satisfying assignment must
satisfy all of the clauses.

We write $\tautology$ to denote both tautology and logical truth.
It arises when a clause
contains both a variable and its complement.
We write $\nil$ to denote logical falsehood.  It is represented by an empty clause.

We make use of the {\em if-then-else} operation, written $\ite$, defined as
$\ite(u,v,w) = (u \booland v) \boolor (\neg u \booland w)$.

When writing clauses, we omit disjunction symbols and use
overlines to denote negation, writing $\neg u\lor v\lor \neg w$ as
$\obar{u}\,v\,\obar{w}$.

\subsection{Resolution Proofs}

Robinson~\cite{robinson-1965} observed that the resolution inference
rule formulated by Davis and
Putnam~\cite{davis60_a_computing_procedure} could form the basis for a
refutation theorem-proving technique for first-order logic.  Here, we
consider its specialization to propositional logic.  For clauses of
the form $C\boolor x$, and $\obar{x} \boolor D$, the resolution rule
derives the new clause $C \lor D$.  This inference is written with a
notation showing the required conditions above a horizontal line, and
the resulting inference (known as the {\em resolvent}) below:
\begin{center}
\def\ScoreOverhang{2pt}
\def\defaultHypSeparation{\hskip 0.1in}
  \begin{prooftree}
    \AxiomC{$C \boolor x$}
    \AxiomC{$\obar{x} \boolor D$}
    \BinaryInfC{$C \boolor D$}
  \end{prooftree}
\end{center}  
Intuitively, the resolution rule is based on the property that implication is
transitive.  To see this, let proposition $p$ denote $\neg C$, and
proposition $q$ denote $D$.  Then $C \boolor x$ is equivalent to
$p \rightarrow x$, $\obar{x} \boolor D$ is equivalent to
$x \rightarrow q$, and $C \boolor D$ is equivalent to $p \rightarrow q$.  In other
words, the resolution rule encodes the property that if $p \rightarrow
x$ and $x \rightarrow q$, then $p \rightarrow q$.  As a special case, when $C$ contains a literal $l$ and $D$ contains its complement $\obar{l}$,
then the resolvent of $C \lor x$ and $D \lor \obar{x}$ will be a tautology.

Resolution provides a mechanism for proving that a set of clauses is
unsatisfiable.  Suppose the input consists of $m$ clauses.
A resolution proof is given as a {\em trace} consisting of a
series of {\em steps} $S$, where each step $s_i$ consists of a clause
$C_i$ and a (possibly empty) list of antecedents $A_i$, where each
antecedent is the index of one of the previous steps.  The first set
of steps, denoted $S_m$, consists of the input clauses without any
antecedents.  Each successive step then consists of a clause and a set
of antecedents, such that the clause can be derived from the
clauses in the antecedents by one or more resolution steps.  It follows by
transitivity that for each step $s_i$, with $i > m$, clause $C_i$ is
logically implied by the input clauses, written $S_m \turnstile C_i$.
If, through a series of steps, we can reach a step $s_t$ where $C_t$ is
the empty clause, then the trace provides a proof
that $S_m \turnstile \nil$, i.e., the set of input clauses is not
satisfiable.



A typical resolution proof contains many applications
of the resolution rule.  These enable deriving sequences of
implications that combine by transitivity.  For example, consider the following
implications, shown both as formulas and as clauses:
\begin{center}
  \begin{tabular}{cc}
    \makebox[15mm]{Formula} & \makebox[15mm]{Clause} \\
    \midrule
    $a \rightarrow b$ & $\obar{a} \, b$ \\
    $x \rightarrow (b \rightarrow c)$ & $\obar{x}\,\obar{b} \, c$ \\
    $c \rightarrow d$ & $\obar{c} \, d$ \\
    \midrule
    $x \rightarrow (a \rightarrow d)$ & $\obar{x}\,\obar{a} \, d$ \\
  \end{tabular}
\end{center}
We can derive the final clause from the first three using two resolution steps:
\begin{center}
\def\ScoreOverhang{2pt}
\def\defaultHypSeparation{\hskip 0.1in}
  \begin{prooftree}
    \AxiomC{$\obar{x}\,\obar{b}\,c$}
    \AxiomC{$\obar{a}\,b$}
    \BinaryInfC{$\obar{x}\,\obar{a}\,c$}
    \AxiomC{$\obar{c}\,d$}
    \BinaryInfC{$\obar{x}\,\obar{a}\,d$}
  \end{prooftree}
\end{center}

\subsection{Reverse Unit Propagation (RUP)}

Reverse unit propagation (RUP) provides an easily checkable way to
express a linear sequence of resolution operations as a single proof
step~\cite{Goldberg,RUP}.  It is the core rule supported by standard
proof checkers~\cite{RAT,wetzler14_drattrim} for propositional logic.
Let $C = \lit_1 \, \lit_2 \, \cdots \, \lit_p$ be a clause to be
proved and let $D_1, D_2, \ldots, D_k$ be a sequence of supporting
{\em antecedent} clauses occurring earlier in the proof.  A RUP step
proves that $\bigwedge_{1\leq i \leq k} D_i \rightarrow C$ by showing
that the combination of the antecedents plus the negation of $C$ leads
to a contradiction.  The negation of $C$ is the formula
$\overline{\lit}_1 \land \overline{\lit}_2 \land \cdots \land
\overline{\lit}_p$ having a CNF representation consisting of $p$ unit
clauses of the form $\obar{\lit}_i$ for $1 \leq i \leq p$.  A RUP
check processes the clauses of the antecedent in sequence, inferring
additional unit clauses.  In processing clause $D_i$, if all but one
of the literals in the clause is the negation of one of the
accumulated unit clauses, then we can add this literal to the
accumulated set.  That is, all but this literal have been falsified,
and so it must be set to true for the clause to be satisfied.  The
final step with clause $D_k$ must cause a contradiction, i.e., all of
its literals are falsified by the accumulated unit clauses.

As an example, consider 
a RUP step to derive $x \rightarrow (a \rightarrow d)$ from the three clauses shown in the earlier example.
A RUP proof would take the following form.  Here, the target and antecedent clauses are listed along the top, while the resulting unit clauses are shown on the bottom, along with the final contradiction.
\begin{center}
  \begin{tabular}{lcccc}
         & \makebox[15mm]{Target}    & \makebox[10mm]{} & \makebox[10mm]{Antecedents} & \makebox[10mm]{} \\
  Clause & $\obar{x}\,\obar{a} \, d$ & $\obar{c} \, d$  & $\obar{x}\,\obar{b} \, c$   & $\obar{a} \, b$ \\
  \midrule
  Units  & $x$, $a$, $\obar{d}$      &  $\obar{c}$             & $\obar{b}$                         & $\nil$ \\
  \end{tabular}
\end{center}
RUP is an alternative formulation of resolution.  For target clause
$C$, it can be seen that applying resolution operations to the
antecedent clauses from right to left will derive a clause $C'$ such
that $C' \subseteq C$.  By {\em subsumption}~\cite{philipp:lpar:2017},
we then have $C' \rightarrow C$.  Compared to listing each resolution
operation as a separate step, using RUP as the basic proof step makes
the proofs more compact.

\subsection{Extended Resolution}

Grigori S. Tseitin~\cite{Tseitin:1983} introduced the
extended-resolution proof framework in a presentation at the Leningrad
Seminar on Mathematical Logic in 1966.  The key idea is to allow the
addition of new {\em extension} variables to a resolution proof in a
manner that preserves the soundness of the proof.  In particular, in
introducing variable $e$, there must be an accompanying set of clauses 
that encode $e \leftrightarrow F$, where $F$ is a formula over
variables (both original and extension) that were introduced earlier.
These are referred to as the {\em defining clauses} for extension
variable $e$.  Variable $e$ then provides a shorthand notation by
which $F$ can be referenced multiple times.  Doing so can reduce the
size of a clausal representation of a problem by an exponential
factor.  

An extension variable $e$ is introduced into the proof by including
its defining clauses in the list of clauses being generated.  The
proof checker must then ensure that the defining clauses obey the
requirements for extension variables, as is discussed below.  Thereafter, other clauses can
include the extension variable or its complement,
and they can list the defining clauses as antecedents.

Tseitin transformations are commonly used to encode a logic circuit or
formula as a set of clauses without requiring the formulas to be
``flattened'' into a conjunctive normal form over the circuit inputs
or formula variables.  These introduced variables are called {\em
  Tseitin variables} and are considered to be part of the input
formula.  An extended resolution proof takes this concept one step
further by introducing additional variables as part of the proof.  The
proof checker must ensure that these {\em extension} variables are
used in a way that does not result in an unsound proof.
Some problems for which the minimum resolution proof must be of
exponential size can be expressed with polynomial-sized proofs in
extended resolution~\cite{Cook:1976}.


\subsection{Clausal Proofs}

We use a {\em clausal proof system} to validate our proofs based on
the DRAT proof framework ~\cite{RAT}.  This framework provides
supports both extended resolution and resolution operations based on a proof
rule that generalizes reverse unit propagation.
There are a number of
fast and formally-verified checkers for these
proofs~\cite{Cruz-FilipeMS17-TACAS,wetzler14_drattrim,Lammich}.
The checker ensures that all
extension variables are used properly and that
each new clause can be derived via RUP from its antecedent
clauses.

As in a resolution proof,
a clausal proof is given as a trace, where each step $s_i$ consists of a clause $C_i$ and a list of antecedents $A_i$, where the initial $m$ clauses are the input clauses.
Let $S_m$ denote the set of input clauses, and for 
$i > m$, define $S_i$ inductively as $S_i = S_{i-1} \cup \{ C_i \}$.
The proof steps $s_{m+1}, \dots, s_t$ represent a derivation from $S_m$ to $S_t$. 
A clausal proof is a refutation if $S_t$ contains the empty clause. 
Step $s_i$ in a proof is valid if the equisatisfiability\footnote{Two Boolean formulas are {\em equisatisfiable} if they are either both satisfiable or both unsatisfiable.} of $S_{i-1}$ and $S_{i}$ can be
checked using a polynomially decidable redundancy property.  
For the case where $C_i$ was obtained via RUP, we can simply perform a RUP check using $C_i$ and the antecedents.
In case $C_i$ is one of the defining clauses for some extension variable $e$,
the checker must ensure that the clause is {\em blocked}~\cite{kullman:dam:1999}.  That is,
all possible resolvents of $C_i$ with clauses in $S_{i-1}$ that contain $e$
must be tautologies. The blocked clause proof system is a generalization
of extended resolution and allows the addition of blocked clauses that are 
blocked on non-extension variables. However, we do not use such capabilities
in our proofs.


Clausal proofs also allow the removal of clauses.  A proof can
indicate that clause $C_j$ can be removed after step $s_i$
if it will not be used as an antecedent
in any step $s_k$ with $k > i$.  With this restriction, clause
deletion does not affect the integrity of the proof.  As 
the experimental results of Section~\ref{sect:experimental} demonstrate, deleting
clauses that are no longer needed can substantially reduce the number
of clauses the checker must track while processing a proof.


\subsection{Binary Decision Diagrams}

Reduced, Ordered Binary Decision Diagrams (which we refer to as simply
``BDDs'') provide a canonical form for representing Boolean functions,
and an associated set of algorithms for constructing them and testing
their properties~\cite{Bryant:1986}.
With BDDs, functions are defined over a set
of variables $X = \{x_1, x_2, \ldots, x_n\}$.
We let $\leafzero$ and $\leafone$ denote the two leaf
nodes, representing the constant functions $\zero$ and $\one$, respectively.

Each nonterminal node $\nodeu$ has an associated variable $\var(\nodeu)$ and
children $\hi(\nodeu)$, indicating the case where the node variable has
value 1, and $\lo(\nodeu)$, indicating the case where the node variable
has value 0.

Two lookup tables---the {\em unique table} and the {\em operation cache}---are critical for guaranteeing the canonicity of the
BDDs and for ensuring polynomial performance of the BDD construction
algorithms.

A node $\nodeu$ is stored in a unique table, indexed by a key of the form
$\langle \var(\nodeu), \hi(\nodeu), \lo(\nodeu)\rangle$, so that isomorphic nodes are never
created.  The nodes are shared across all of the BDDs~\cite{minato-dac-1990}.  In presenting
algorithms, we assume a function $\opname{GetNode}(\var(\nodeu), \hi(\nodeu),\lo(\nodeu))$
 that checks the unique table and either returns the node
stored there, or it creates a new node and enters it into the table.
With this table, we can guarantee that the subgraphs with root nodes
$\nodeu$ and $\nodev$ represent the same Boolean function if and only if $\nodeu=\nodev$.
We can therefore uniquely identify Boolean functions with their BDD root nodes.

\begin{figure}
\begin{tabbing}
xxxxxxx\=xxxxxxx\=xxxxxxx\=xxxxxxx\=\kill
\applyop(\op, $\nodeu_1$, \ldots, $\nodeu_k$) \\
\>\keyif{} \opname{IsTerminal}($\op$, $\nodeu_1$, \ldots, $\nodeu_k$): \\
\>\>\keyreturn{} \opname{TerminalValue}($\op$, $\nodeu_1$, \ldots, $\nodeu_k$) \\
\>$K \assign \langle \op, \nodeu_1, \ldots, \nodeu_k \rangle$ \\
\>\keyif{} $K \in \cache$: \\
\>\>\keyreturn{} $\cache[K]$ \\
\>$\nodew \assign{} \;$ \opname{ApplyRecur}($\op$, $\nodeu_1$, \ldots, $\nodeu_k$) \\
\>$\cache[K] \assign{} \nodew$ \\
\>\keyreturn{} $\nodew$
\end{tabbing}
\caption{General structure of the Apply algorithm.  The operation for a specific logical operation $\op$ is determined by its terminal cases and its recursive structure.}
\label{fig:apply}
\end{figure}  

BDD packages support multiple operations for constructing and testing
the properties of Boolean functions represented by BDDs.  A number of
these are based on the {\em Apply} algorithm~\cite{Bryant:1986}.
Given a set of BDD roots $\nodeu_1, \nodeu_2, \ldots, \nodeu_k$ representing
functions $f_1, f_2, \ldots, f_k$ respectively, and a Boolean operation
$\op$, the algorithm generates the BDD
representation $\nodew$ of the operation applied to those functions.  For example, with $k = 2$, and $\op = \andop$,
$\applyop(\andop, \nodeu_1, \nodeu_2)$ returns the root node for the BDD representation of $f_1 \land f_2$.

Figure \ref{fig:apply} shows pseudo-code describing the overall
structure of the Apply algorithm.  The details for a specific
operation are embodied in the functions \opname{IsTerminal},
\opname{TerminalValue}, and \opname{ApplyRecur}.  The first two of
these detect terminal cases and what value to return when a
terminal case is encountered.  The third describes how to handle the
general case, where the arguments must be expanded recursively.  The
algorithm makes use of {\em memoizing}, where previously computed
results are stored in an operation cache, indexed by a key consisting of the
operands~\cite{michie:1968}.  Whenever possible, results are retrieved from this cache, avoiding the need to perform redundant calls to \opname{ApplyRecur}.
With this cache, the worst
case number of recursive steps required by the algorithm is bounded by
the product of the sizes (in nodes) of the arguments.

\section{Proof Generation During BDD Construction}
\label{sect:proof}

In our formulation, every newly created BDD node $\nodeu$ is assigned
an extension variable $u$.
(As notation, nodes are denoted by boldface
characters, possibly with subscripts, e.g., $\nodeu$, $\nodev$, and $\nodev_1$, while their corresponding extension
variables are denoted with a normal face, e.g., $u$, $v$,
and $v_1$.)
We then
extend the Apply algorithm to generate proofs based on the recursive
structure of the BDD operations.

Let $S_m$ denote the set of input clauses.  Our
goal is to generate a proof that $S_m \turnstile \nil$, i.e., there is
no satisfying assignment for these clauses.  Our BDD-based approach
generates a sequence of BDDs with root nodes
$\nodeu_1, \nodeu_2, \ldots, \nodeu_t$, where $\nodeu_t = \leafzero$,
based on a combination of the following operations.
(The exact sequencing of operations is determined
by the {\em evaluation mechanism}, as is described in Section~\ref{sect:experimental}.)
\begin{enumerate}
\item
  For input clause $C_i$ generate its BDD representation $\nodeu_i$ using a series of Apply operations to perform the disjunctions.
\item
  For roots $\nodeu_j$ and $\nodeu_k$, generate the BDD representation of their conjunction $\nodeu_l = \nodeu_j \booland \nodeu_k$ using the Apply operation to perform conjunction.
\item
  For root $\nodeu_j$ and some set of variables $Y \subseteq X$, perform existential quantification: $\nodeu_k = \exists Y \, \nodeu_j$.
\end{enumerate}  

Although the existential quantification operation is not mandatory for a
BDD-based SAT solver, it can greatly improve its performance~\cite{franco-sat-2004}.
It is the BDD counterpart to
Davis-Putnam variable elimination on
clauses~\cite{davis60_a_computing_procedure}.  As the notation
indicates, there are often multiple variables that can be eliminated
simultaneously.  Although the operation can cause a BDD to increase in
size, it generally causes a reduction.  Our experimental results
demonstrate the importance of this operation.

As these operations proceed, we simultaneously generate a set of proof
steps.  The details of each step are given later in the presentation.
For each BDD generated, we maintain the proof invariant that the extension variable $u_j$ associated with root node $\nodeu_j$ satisfies $S_m \turnstile u_j$.
\begin{enumerate}
\item Following the generation of the BDD $\nodeu_i$ for input clause $C_i$, we also generate a proof that $C_i \turnstile u_i$.
  This is described in Section~\ref{sect:clauses}.
\item Justifying the results of conjunctions requires two parts:
\begin{enumerate}
\item
Using a modified version of the Apply algorithm for conjunction we follow the
structure of its recursive calls to generate a proof that the algorithm
preserves implication: $u_j \land u_k \rightarrow u_l$.  This is
described in Section~\ref{sect:apply:and}.
\item 
This implication can be combined with the earlier proofs that $S_m \turnstile u_j$ and $S_m \turnstile u_k$ to prove $S_m \turnstile u_l$.
\end{enumerate}
\item Justifying the quantification also requires two parts:
\begin{enumerate}
\item Following the generation of $\nodeu_k$ via existential
  quantification, we perform a separate check that their associated extension variables satisfy
  $u_j \rightarrow  u_k$.  This check uses a proof-generating version of the Apply
  algorithm for implication checking.
  This is described in Section~\ref{sect:apply:imply}.
\item
This implication can be combined with the earlier proof that $S_m \turnstile u_j$ to prove $S_m \turnstile u_k$.
\end{enumerate}
\end{enumerate}  

Compared to the prior work by Sinz and Biere~\cite{ebddres}, our key
refinement is to handle arbitrary existential quantification
operations.  (When implementing a SAT solver, these quantifications
must be applied in restricted ways \cite{weaver_jsat_2006}, but since
proofs of unsatisfiability only require proving implication, we need
not be concerned with the details of these restrictions.)  Rather than attempting to
track the detailed logic underlying the quantification operation, we
run a separate check that implication is preserved.  As is the case
with many BDD packages, our implementation can perform existential
quantification of an arbitrary set of variables in a single pass
over the argument BDD\@.  We only need to perform a single
implication check for the entire quantification.

Sinz and Biere's construction assumed there were special extension
variables $n_1$ and $n_0$ to represent the BDD leaves $\leafone$ and
$\leafzero$.  Their proofs then includes unit clauses $n_1$ and
$\obar{n}_0$ to force these variables to be set to 1 and 0, respectively.
We have found that these special variables are not required and instead
directly associate leaves $\leafone$ and
$\leafzero$ with $\tautology$ and $\nil$, respectively.

The $n$ variables in the input clauses all have associated BDD
variables.  The proof then introduces an extension variable $u$ every time
a new BDD node $\nodeu$ is created.  In the actual implementation, the
extension variable (an integer) is stored as one of the fields in the
node representation.

When creating a new node, the \opname{GetNode} function adds (up to) four defining clauses for the
associated extension variable.
For node $\nodeu$ with variable $\var(\nodeu) = x$, $\hi(\nodeu) = \nodeu_1$, and
  $\lo(\nodeu) = \nodeu_0$, the clauses are:
  \begin{center}
    \begin{tabular}{ccc}
\toprule
      Notation & Formula & Clause \\
\midrule
      $\stephd(\nodeu)$ & $x \rightarrow (u \rightarrow u_1)$ & $\obar{x} \, \obar{u} \, u_1$ \\
      $\stepld(\nodeu)$ & $\obar{x} \rightarrow (u \rightarrow u_0)$ & $x \, \obar{u} \, u_0$ \\
      $\stephu(\nodeu)$ & $x \rightarrow(u_1 \rightarrow u)$ & $\obar{x} \, \obar{u}_1 \, u$ \\
      $\steplu(\nodeu)$ & $\obar{x} \rightarrow (u_0 \rightarrow u)$ & $x \, \obar{u}_0 \, u$ \\
\bottomrule
    \end{tabular}
  \end{center}
The names for these clauses combines an indication of whether they
correspond to variable $x$ being 1 (\textsf{H}) or 0 (\textsf{L}) and
whether they form an implication from the node down to its child
(\textsf{D}) or from the child up to its parent (\textsf{U}).  When
one of the child nodes $\nodeu_0$ or $\nodeu_1$ is a leaf, some of these
defining clauses will degenerate into tautologies, and some will
reduce to just two literals.  Tautologies are not included in the proof.
These defining clauses encode the assertion:
\begin{eqnarray*}
u & \leftrightarrow & \ite(x, u_1, u_0)
\end{eqnarray*}  
satisfying Tseitin's restriction on the use of extension variables.
Each clause is
numbered according to its step number in the trace.  

\subsection{Generating BDD Representations of Clauses}
\label{sect:clauses}

\begin{figure}
\begin{center}
\setlength{\unitlength}{10pt}
\begin{picture}(6.5,14)
  \put(0,29){a.~BDD representation}
  \put(2,13){\circle{2.0}}
  \put(3,9){\circle{2.0}}
  \put(4,5){\circle{2.0}}
  \put(0,0){\framebox(2,2){1}}
  \put(4,0){\framebox(2,2){0}}
  \put(2,13){\clabel{$a$}}
  \put(3,9){\clabel{$b$}}
  \put(4,5){\clabel{$c$}}      
  \put(3.5,13){\llabel{$\nodeu_a$}}
  \put(4.5,9){\llabel{$\nodeu_b$}}
  \put(5.5,5){\llabel{$\nodeu_c$}}  
  \put(1.08,2){\line(1,12){0.834}}
  \multiput(1.25,2.1)(0.1,0.4){15}{\line(1,4){0.06}}
  \put(1.75,2.0){\line(3,4){1.65}}
  \multiput(2.76,9.98)(-0.1,0.4){5}{\line(-1,4){0.06}}  
  \put(3.76,5.96){\line(-1,4){0.52}}
  \multiput(4.76,2.1)(-0.1,0.4){5}{\line(-1,4){0.06}}
\end{picture}
\end{center}
\medskip
\begin{center}
  \begin{tabular}{lccccccc}
   & \makebox[10mm]{Target} & \makebox[10mm]{$\stephu(\nodeu_a)$} & \makebox[10mm]{$\steplu(\nodeu_a)$} & \makebox[10mm]{$\steplu(\nodeu_b)$} & \makebox[10mm]{$\stephu(\nodeu_b)$} & \makebox[10mm]{$\stephu(\nodeu_c)$} & \makebox[10mm]{$C$}  \\
  Clause & $u_a$            & $\obar{a}\, u_a$        & $a \, \obar{u}_b \, u_a$        & $b \, u_b$                            & $\obar{b} \, \obar{u}_c \, u_b$ & $\obar{c} \, u_c$             & $a\, \obar{b} \, c$ \\
  \midrule
  Units & $\obar{u}_a$      & $\obar{a}$              & $ \obar{u}_b$                   & $b$                                   & $\obar{u}_c$                    & $\obar{c}$                    & $\nil$ \\
  \end{tabular}
\end{center}


\caption{BDD representation of clause $C = a\,\obar{b}\,c$ and the justification of root unit clause $\nodeu_a$ with one RUP step.}
\label{fig:resolution:clause}
\end{figure}  

The BDD representation for a clause $C$ has a simple, linear
structure.  For root node $\nodeu$, it is easy to prove that
$C \turnstile u$ using one RUP step.  The general algorithm is described by Sinz and
Biere~\cite{ebddres}.  Here we illustrate the idea via an example.

Figure \ref{fig:resolution:clause} shows the BDD representation of
clause $C = a\,\obar{b}\,c$.  As can be seen, the BDD for a clause has
a very specific structure.  For each literal in the clause, there is a
node labeled by the variable, with one child being leaf $\leafone$ and
the other being either the node for the next literal in the variable
ordering or leaf $\leafzero$.  The lower part of the figure
shows a RUP justification of $C \turnstile u_a$, where $\nodeu_a$ is the
root node of the BDD.  The proof uses the antecedents $\stephu(\nodeu)$
and $\steplu(\nodeu)$ for each node $\nodeu$ in the BDD (except for the
tautological case representing the final edge to $\leafzero$), with
the final antecedent being the clause itself.  The
RUP steps introduce the complements of the clause variables as unit clauses, causing a contradiction with the input clause.
The order in which the two defining clauses for a node are listed in the antecedent
depends on whether the variable is positive or negative in the clause.
As this example demonstrates, 
we can generate a single proof step for $C_i \turnstile u_i$
for each input clause $C_i$.

\subsection{Performing Conjunctions}
\label{sect:apply:and}

\begin{figure}
\adjustbox{valign=t}{
\begin{minipage}{1.5in}
  \begin{tabular}{cc}
    \toprule
    \multicolumn{2}{c}{Terminal Cases} \\
    \makebox[15mm]{Condition} & \makebox[15mm]{Result} \\
    \midrule
    $\nodeu = \nodev$ & ($\nodeu$, $\tautology$) \\
    $\nodeu = \leafzero$ & ($\leafzero$, $\tautology$) \\
    $\nodev = \leafzero$ & ($\leafzero$, $\tautology$) \\    
    $\nodeu = \leafone$ & ($\nodev$, $\tautology$) \\
    $\nodev = \leafone$ & ($\nodeu$, $\tautology$) \\   
    \bottomrule 
  \end{tabular}  
\end{minipage}}
\begin{minipage}[t]{3in}
\begin{tabbing}
xxxxxxx\=xxxxxxx\=xxxxxxx\=xxxxxxx\=\kill
\opname{ApplyRecur}(\andop, $\nodeu$, $\nodev$) \\
\>$J \assign{} \{\}$ \>\> \\
\>$x \assign{} \min(\var(\nodeu), \var(\nodev))$ \\
\>\keyif{} $x = \var(\nodeu)$:\\
\>\>$\nodeu_1, \nodeu_0 \assign{} \hi(\nodeu), \lo(\nodeu)$\\
\>\>$J \assign J \cup \{ \stephd(\nodeu), \stepld(\nodeu) \}$ \\
\>\keyelse:\>$\nodeu_1, \nodeu_0 \assign{} \nodeu, \nodeu$\\
\>\keyif{} $x = \var(\nodev)$:\\
\>\>$\nodev_1, \nodev_0 \assign{} \hi(\nodev), \lo(\nodev)$\\
\>\>$J \assign J \cup \{ \stephd(\nodev), \stepld(\nodev) \}$ \\
\>\keyelse:\>$\nodev_1, \nodev_0 \assign{} \nodev, \nodev$\\
\>$\nodew_1, s_1 \assign{} \applyop(\andop, \nodeu_1, \nodev_1)$\\
\>$\nodew_0, s_0 \assign{} \applyop(\andop, \nodeu_0, \nodev_0)$\\
\>$J \assign J \cup \{ s_1, s_0 \}$ \\
\>\keyif{} $\nodew_1 = \nodew_0$:\\
\>\>$\nodew \assign{} \nodew_1$\\
\>\keyelse{}:\\
\>\>$\nodew \assign \opname{GetNode}(x, \nodew_1, \nodew_0$)\\
\>\>$J \assign J \cup \{ \stephu(\nodew), \steplu(\nodew) \}$ \\
\>$s \assign \opname{JustifyAnd}(\langle \nodeu, \nodev, \nodew \rangle, J)$ \\
\>\keyreturn{} $(\nodew, s)$ 
\end{tabbing}
\end{minipage}
\caption{Terminal cases and recursive step
of the Apply operation for conjunction, modified for proof generation.  Each call returns both a node and a proof step.}
\label{fig:apply:and}
\end{figure}  


The key idea in generating proofs for the conjunction operation is to
follow the recursive structure of the Apply algorithm.  We do this by
integrating proof generation into the Apply procedures, as is
shown in Figure \ref{fig:apply:and}.  This follows the standard form of the Apply algorithm (Figure \ref{fig:apply}),
with the novel feature that each result includes both a BDD node $\nodew$ and a proof step number $s$.
For arguments $\nodeu$ and $\nodev$, step
$s$ lists clause $\obar{u}\,\obar{v}\,w$ along with antecedents defining
a RUP proof of the implication $u \land v \rightarrow w$.

As the table of terminal cases shows, these cases all correspond
to tautologies.  For example, the case of $\nodeu = \leafone$, giving $\nodew = \nodev$ is
justified by that tautology $\tautology \booland v \rightarrow v$.
Failing a terminal or previously computed case, the function must recurse, branching on the
variable $x$ that is the minimum of the two root variables.  The
procedure accumulates a set of proof steps $J$ to
be used in the implication proof.  These include the two steps
(possibly tautologies) from the two recursive calls.  At the end, it
invokes a function \opname{JustifyAnd} to generate the
required proof.  In returning the pair $(\nodew, s)$, this value will be stored in the operation cache and returned as the result of the Apply operation.

\begin{figure}
\begin{center}
\begin{tabular}{ccc}
\toprule
  Label & Formula & Clause \\
\midrule
  \cuhd & $\stephd(\nodeu)$ & $\obar{x}\,\obar{u}\,u_1$ \\
  \culd & $\stepld(\nodeu)$ & ${x}\,\obar{u}\,u_0$ \\
  \cvhd & $\stephd(\nodev)$ & $\obar{x}\,\obar{v}\,v_1$ \\
  \cvld & $\stepld(\nodev)$ & ${x}\,\obar{v}\,v_0$ \\
  \cwhu & $\stephu(\nodew)$ & $\obar{x}\,\obar{w}_1\,w$ \\
  \cwlu & $\steplu(\nodew)$ & ${x}\,\obar{w}_0\,w$ \\
  \candh & $u_1 \booland v_1 \rightarrow w_1$ & $\obar{u}_1\,\obar{v}_1\,w_1$ \\
  \candl & $u_0 \booland v_0 \rightarrow w_0$ & $\obar{u}_0\,\obar{v}_0\,w_0$ \\
\bottomrule
\end{tabular}  
\end{center}  
\caption{Supporting clauses for standard
step of the Apply algorithm for conjunction operations}
\label{fig:clause:and:standard}
\end{figure}

\begin{figure}
\def\ScoreOverhang{2pt}
\def\defaultHypSeparation{\hskip 0.15in}

\begin{center}
  \begin{tabular}{lcccccc}
   & \makebox[15mm]{Target} & \makebox[10mm]{} & \makebox[10mm]{$\cuhd$} & \makebox[10mm]{$\cvhd$} & \makebox[10mm]{$\cwhu$} & \makebox[10mm]{$\candh$}  \\
  Clause & $\obar{x} \, \obar{u}\,\obar{v} \, w$ & & $\obar{x} \, \obar{u} \, u_1$ & $\obar{x} \, \obar{v} \, v_1$ & $\obar{x} \, \obar{w}_1 \, w$ & $\obar{u}_1\,\obar{v}_1 \, w_1$ \\
  \midrule
  Units & $x$, $u$, $v$, $\obar{w}$ & & $u_1$ & $v_1$ & $\obar{w}_1$ & $\nil$ \\
\\
  & \makebox[15mm]{Target} & \makebox[10mm]{Previous} & \makebox[10mm]{$\culd$} & \makebox[10mm]{$\cvld$} & \makebox[10mm]{$\cwlu$} & \makebox[10mm]{$\candl$}  \\
  Clause & $\obar{u}\,\obar{v} \, w$ & $\obar{x} \, \obar{u}\,\obar{v} \, w$ & $x \, \obar{u} \, u_0$ & $x \, \obar{v} \, v_0$ & $x \, \obar{w}_0 \, w$ & $\obar{u}_0\,\obar{v}_0 \, w_0$ \\
  \midrule
  Units & $u$, $v$, $\obar{w}$ & $\obar{x}$ & $u_0$ & $v_0$ & $\obar{w}_0$ & $\nil$ \\
  \end{tabular}
\end{center}

\caption{RUP proof steps for standard recursive step of the conjunction operation}
\label{fig:resolution:and:standard}
\end{figure}

\subsubsection{Proof Generation for the Standard Case}

A proof generated by \applyop{} with operation $\andop$ inducts
on the structure of the argument and result BDDs.
That is, it assumes that the result nodes $\nodew_1$ and $\nodew_0$ of the
recursive calls to arguments $\nodeu_1$ and $\nodev_1$ and to $\nodeu_0$ and $\nodev_0$
satisfy the implications $u_1 \land v_1 \rightarrow w_1$ and $u_0 \land
v_0 \rightarrow w_0$, and that these calls generated proof steps $s_1$
and $s_0$ justifying these implications.  For the standard case, where
none of the equalities hold and the recursive calls do not yield
tautologies, the supporting clauses for the proof are shown in
Figure \ref{fig:clause:and:standard}.  That is the set $J$ contains
references to eight clauses, which we identify by labels.  Six of these are defining clauses:
the downward clauses for the argument nodes (labeled \cuhd, \cvhd, \culd, and \cvld)
and the upward clauses for the result (labeled \cwhu{} and \cwlu).
The other two are implications for the two recursive calls, labeled (\candh{} and \candl).
We partition these supporting clauses into two sets:
\begin{eqnarray}
  \andchain_H & = &  \cuhd{}, \cvhd{}, \cwhu{}, \candh{} \label{eqn:achainh} \\
  \andchain_L & = &  \culd{}, \cvld{}, \cwlu{}, \candl{} \label{eqn:achainl} 
\end{eqnarray}

These supporting clauses are used to derive the target clause
$u \land v \rightarrow w$ using the two RUP steps shown in Figure
\ref{fig:resolution:and:standard}.  The first step proves the weaker
target $x \rightarrow (u \land v \rightarrow w)$, having clausal
representation $\obar{x}\,\obar{u}\,\obar{v}\,w$ using the supporting clauses in $\andchain_H$.
The second step proves the full target, having clausal representation $\obar{u}\,\obar{v}\,w$.  It uses both the weaker result plus the supporting clauses in $\andchain_L$.

\subsubsection{Proof Generation for Special Cases}

\begin{figure}
\begin{minipage}{\textwidth}
A)  $\nodeu_1 = \nodeu_0$ and $\nodew_1 = \nodew_0$
\end{minipage}
\begin{center}
  \begin{tabular}{lcccccc}
         & \makebox[15mm]{Target}                & \makebox[10mm]{} & \makebox[10mm]{}              & \makebox[10mm]{$\cvhd$}       & \makebox[10mm]{}              & \makebox[10mm]{$\candh$}  \\
  Clause & $\obar{x} \, \obar{u}\,\obar{v} \, w$ &                  &                               & $\obar{x} \, \obar{v} \, v_1$ &                               & $\obar{u}\,\obar{v}_1 \, w$ \\
  \midrule
  Units  & $x$, $u$, $v$, $\obar{w}$             &                  &                               & $v_1$                         &                               & $\nil$ \\
\\
         & \makebox[15mm]{Target}                & \makebox[10mm]{Previous}              & \makebox[10mm]{}        & \makebox[10mm]{$\cvld$} & \makebox[10mm]{}        & \makebox[10mm]{$\candl$}  \\
  Clause & $\obar{u}\,\obar{v} \, w$             & $\obar{x} \, \obar{u}\,\obar{v} \, w$ &                         & $x \, \obar{v} \, v_0$  &                         & $\obar{u}\,\obar{v}_0 \, w$ \\
  \midrule
  Units  & $u$, $v$, $\obar{w}$                  & $\obar{x}$                            &                         & $v_0$                   &                         & $\nil$ \\
  \end{tabular}
\end{center}
\begin{minipage}{\textwidth}
B) $\nodeu_1 = \leafone$
\end{minipage}
\begin{center}
  \begin{tabular}{lcccccc}
         & \makebox[15mm]{Target}                 & \makebox[10mm]{}  & \makebox[10mm]{} & \makebox[10mm]{$\cvhd$}       & \makebox[10mm]{$\cwhu$}       & \makebox[10mm]{}  \\
  Clause & $\obar{x} \, \obar{u}\, \obar{v} \, w$ &                   &                  & $\obar{x} \, \obar{v} \, v_1$ & $\obar{x} \, \obar{v}_1 \, w$ &                   \\
  \cmidrule{1-6}
  Units & $x$, $u$, $v$, $\obar{w}$               &                   &                  & $v_1$                         & $\nil$ &  \\
\\
         & \makebox[15mm]{Target}      & \makebox[10mm]{Previous}             & \makebox[10mm]{$\culd$} & \makebox[10mm]{$\cvld$} & \makebox[10mm]{$\cwlu$} & \makebox[10mm]{$\candl$}  \\
  Clause & $\obar{u} \, \obar{v} \, w$ & $\obar{x} \obar{u} \, \obar{v} \, w$ & $x\,\obar{u}\,u_0$      & $x \, \obar{v} \, v_0$  & $x \, \obar{w}_0 \, w$  & $\obar{u}_0 \, \obar{v}_0 \, w_0$ \\
  \midrule
   Units & $u$, $v$, $\obar{w}$        & $\obar{x}$                           & $u_0$                   & $v_0$                   & $\obar{w}_0$            & $\nil$ \\
  \end{tabular}
\end{center}
\begin{minipage}{\textwidth}
C)  $\nodeu_1 = \leafzero$
\end{minipage}
\begin{center}
  \begin{tabular}{lcccccc}
  & \makebox[15mm]{Target} & \makebox[10mm]{$\cuhd$} & \makebox[10mm]{$\culd$} & \makebox[10mm]{$\cvld$} & \makebox[10mm]{$\cwlu$} & \makebox[10mm]{$\candl$}  \\
  Clause & $\obar{u}\,\obar{v} \, w$ & $\obar{x} \, \obar{u}$ & $x \, \obar{u} \, u_0$ & $x \, \obar{v} \, v_0$ & $x \, \obar{w}_0 \, w$ & $\obar{u}_0\,\obar{v}_0 \, w_0$ \\
  \midrule
  Units & $u$, $v$, $\obar{w}$ & $\obar{x}$ & $u_0$ & $v_0$ & $\obar{w}_0$ & $\nil$ \\
  \end{tabular}
\end{center}
\caption{RUP proof steps for conjunction for illustrative special cases}
\label{fig:resolution:and:special}
\end{figure}  

The proof structure shown in Figure \ref{fig:resolution:and:standard}
only holds for the standard form of the recursion.  However, there
are many special cases, such as when a recursive call yield
a tautologous result, when some of the child nodes are equal, and when
the two recursive calls return the same node.  Fortunately, a general
approach can handle the many special cases that arise.
The examples shown in Figure~\ref{fig:resolution:and:special} illustrate a range of
possibilities.  Based on these and the standard case of
Figure~\ref{fig:resolution:and:standard},
we will show how to handle all of the cases with a simple algorithm.

Figure~\ref{fig:resolution:and:special}A illustrates the case where
some of the nodes in the recursive calls are equal.  In particular,
when $\var(\nodeu) > \var(\nodev)$, the recursion will split with $\nodeu_1 =
\nodeu_0 = \nodeu$.  This will cause supporting clauses $\cuhd$ and
$\culd$ to be tautologies.  This example also has $\nodew_1 = \nodew_0
= \nodew$, as will occur when the two recursive calls return identical
result.  This will cause supporting clauses $\cwhu$ and $\cwlu$ to be
tautologies.  The two sets of equalities will cause supporting clause
$\candh$ to be $\obar{u}\,\obar{v}_1\,\obar{w}$ and supporting clause
$\candl$ to be $\obar{u}\,\obar{v}_0\,\obar{w}$.  As can be seen, the
resulting proof will consist of the same two steps as the standard
form, but with fewer supporting clauses.

Figure~\ref{fig:resolution:and:special}B illustrates the case where
$\nodeu_1=\leafone$, and therefore the first recursive call generates a
tautologous result.  This case will cause $\nodew_1 = \nodev_1$, and therefore
supporting clause $\cwhu$ will be $\obar{x}\,\obar{v}_1\,w$.  In
addition, supporting clauses $\cuhd$ and $\candh$ will be tautologies.
Despite these changes, the proof will still have the same two-step
structure as the standard case.

Finally, Figure~\ref{fig:resolution:and:special}C illustrates the case
where $\nodeu_1=\leafzero$, and therefore the first recursive call again
generates a tautologous result.  This case will cause
$\nodew_1 =\leafzero$, and only two clauses among those in $\andchain_H$ will
not be tautogies: $\cuhd$ will be $\obar{x}\,\obar{u}$, and $\cvhd$
will be $\obar{x}\,\obar{v}\,v_1$.  As can be seen, the proof for this
case consists of a single RUP step.  Furthermore, it does not make use
of supporting clause $\cvhd$.

These three examples illustrate the following general properties:
\begin{itemize}
\item When neither $\candh$ nor $\candl$ is a tautology, the proof requires two steps.  Some of the supporting clauses may be tautologies, but the proof can follow
the standard form shown in
in Figure~\ref{fig:resolution:and:standard}.
\item
  When either $\candh$ or $\candl$ is a tautology, it may be possible
  to generate a single-step proof.  Otherwise, it can follow the
  standard, two-step form.
\end{itemize}
Given these possibility, our implementation of \opname{JustifyAnd} uses the following strategy:
\begin{enumerate}
\item If supporting clause  $\candh$
  is a tautology, then attempt a
  single-step proof, using the non-tautologous clauses in
  $\andchain_H$ followed by those in $\andchain_L$.
  If this fails, then perform a two-step proof.
\item Similarly, if supporting clause  $\candl$
  is a tautology, then attempt a
  single-step proof, using the non-tautologous clauses in
  $\andchain_L$ followed by those in $\andchain_H$.
  If this fails, then perform a two-step proof.
\item A two-step proof proceeds by first proving the weaker clause
  $\obar{x}\,\obar{u}\,\obar{v}\,w$ using the non-tautologous clauses in $\andchain_H$.  It then uses this result, plus the clauses in $\andchain_L$ to justify
  target clause $\obar{u}\,\obar{v}\,w$.
\end{enumerate}
In all cases, the antecedent is generated by stepping through the
clauses in their specified order, adding only those that cause unit
propagation or conflict.

\subsection{Checking Implication}
\label{sect:apply:imply}

\begin{figure}
\adjustbox{valign=t}{
\begin{minipage}{1.75in}
  \begin{tabular}{cc}
    \toprule
    \multicolumn{2}{c}{Terminal Cases} \\
    \makebox[15mm]{Condition} & \makebox[15mm]{Result} \\
    \midrule
    $\nodeu = \nodev$ & $\tautology$ \\
    $\nodeu = \leafzero$ & $\tautology$ \\
    $\nodev = \leafone$ & $\tautology$ \\    
    $\nodeu = \leafone, \nodev \not = \leafone$ & Error \\
    $\nodev = \leafzero, \nodeu \not = \leafzero$ & Error \\   
    \bottomrule 
  \end{tabular}  
\end{minipage}}
\begin{minipage}[t]{3in}
\begin{tabbing}
xxxxxxx\=xxxxxxx\=xxxxxxx\=xxxxxxx\=\kill
\opname{ApplyRecur}(\opname{Imply}, $\nodeu$, $\nodev$) \\
\>$J \assign{} \{\}$ \\
\>$x \assign{} \min(\var(\nodeu), \var(\nodev))$ \\
\>\keyif{} $x = \var(\nodeu)$:\\
\>\>$\nodeu_1, \nodeu_0 \assign{} \hi(\nodeu), \lo(\nodeu)$\\
\>\>$J \assign J \cup \{ \stephd(\nodeu), \stepld(\nodeu) \}$ \\
\>\keyelse:
\>$\nodeu_1, \nodeu_0 \assign{} \nodeu, \nodeu$\\
\>\keyif{} $x = \var(\nodev)$:\\
\>\>$\nodev_1, \nodev_0 \assign{} \hi(\nodev), \lo(\nodev)$\\
\>\>$J \assign J \cup \{ \stephu(\nodev), \steplu(\nodev) \}$ \\
\>\keyelse:
\>$\nodev_1, \nodev_0 \assign{} \nodev, \nodev$\\
\>$s_1 \assign{} \applyop(\opname{Imply}, \nodeu_1, \nodev_1)$\\
\>$s_0 \assign{} \applyop(\opname{Imply}, \nodeu_0, \nodev_0)$\\
\>$J \assign J \cup \{ s_1, s_0 \}$ \\
\>$s \assign \opname{JustifyImplication}(\langle \nodeu, \nodev \rangle, J)$ \\
\>\keyreturn{} $s$ \\
\end{tabbing}
\end{minipage}
\caption{Terminal cases and recursive step of the Apply algorithm for implication checking}
\label{fig:apply:imply}
\end{figure}

\begin{figure}
\begin{center}
\begin{tabular}{ccc}
\toprule
  Label & Formula & Clause \\
\midrule
  \cuhd & $\stephd(\nodeu)$ & $\obar{x}\,\obar{u}\,u_1$ \\
  \culd & $\stepld(\nodeu)$ & ${x}\,\obar{u}\,u_0$ \\
  \cvhu & $\stephu(\nodev)$ & $\obar{x}\,\obar{v}_1\,v$ \\
  \cvlu & $\steplu(\nodev)$ & ${x}\,\obar{v}_0\,v$ \\
  \cimh & $u_1 \rightarrow v_1$ & $\obar{u}_1\,v_1$ \\
  \ciml & $u_0 \rightarrow v_0$ & $\obar{u}_0\,v_0$ \\
\bottomrule
\end{tabular}  
\end{center}  
\caption{Clause structure for standard step of implication checking}
\label{fig:clause:imply:standard}
\end{figure}  


\begin{figure}
  \begin{center}
  \begin{tabular}{lccccc}
         & \makebox[15mm]{Target}      & \makebox[10mm]{} & \makebox[10mm]{$\cuhd$}       & \makebox[10mm]{$\cvhu$}       & \makebox[10mm]{$\cimh$}  \\
  Clause & $\obar{x} \, \obar{u} \, v$ &                  & $\obar{x} \, \obar{u} \, u_1$ & $\obar{x} \, \obar{v}_1 \, v$ & $\obar{u}_1 \, v_1$ \\
  \midrule
  Units  & $x$, $u$, $\obar{v}$        &                  & $u_1$                         & $\obar{v}_1$                  & $\nil$ \\
\\
         & \makebox[15mm]{Target} & \makebox[10mm]{Previous}    & \makebox[10mm]{$\culd$} & \makebox[10mm]{$\cvlu$} & \makebox[10mm]{$\ciml$}  \\
  Clause & $\obar{u} \, v$        & $\obar{x} \, \obar{u} \, v$ & $x \, \obar{u} \, u_0$  & $x \, \obar{v}_0 \, v$  & $\obar{u}_0 \, v_0$ \\
  \midrule
  Units  & $u$, $\obar{v}$        & $\obar{x}$                  & $u_0$                   & $\obar{v}_0$            & $\nil$ \\
  \end{tabular}
  \end{center}
\caption{RUP proof steps for standard recursive check of implication checking}
\label{fig:resolution:imply:standard}
\end{figure}  

As described in Section~\ref{sect:proof}, we need not track the
detailed logic of the algorithm that performs existential quantification.
Instead, when the quantification operation applied to
node $\nodeu$ generates node $\nodev$, we generate a proof of implication
afterwards, using the Apply algorithm adapted for implication checking, as shown in 
Figure~\ref{fig:apply:imply}.  A failure of this implication check
would indicate an error in the BDD package, and so its only purpose is
to generate a proof that the implication holds, signaling a fatal error if the
implication does not hold.

This particular operation does not 
generate any new nodes, and so the returned result is simply a proof step number.
The (successful) terminal cases correspond to the tautological cases
$u \rightarrow u$, $\nil \rightarrow v$, and $u \rightarrow \tautology$.

Each recursive step accumulates up to six proof steps as the set $J$
to be used in the implication proof.
Figure \ref{fig:clause:imply:standard} shows the structure of these
clauses for the standard case where neither equality holds, and neither
recursive call returns $\tautology$.
The clauses consist of the two downward
defining clauses for argument $\nodeu$, labeled $\cuhd$ and $\culd$,
the two upward defining clauses for argument $\nodev$, labeled $\cvhu$
and $\cvlu$,
and the
clauses returned by the recursive calls, labeled $\cimh$ and $\ciml$.

Figure \ref{fig:resolution:imply:standard} shows the two RUP steps
required to prove the standard case.  The first step proves the weaker
target $x \rightarrow (u \rightarrow v)$, having clausal
representation $\obar{x}\,\obar{u}\,v$ using the three supporting
clauses containing $\obar{x}$.  The second proves the full target,
having clausal representation $\obar{u}\,v$ using the weaker result
plus the supporting clauses containing $x$.

As with the conjunction operation, there can be many special cases,
but they can be handled with the same general strategy.  If either
recursive result $\cimh$ or $\ciml$ is a tautology, a one-step proof
is attempted.  If that fails, or if neither recursive result is a
tautology, a two-step proof is generated.

\section{Implementation}
\label{sect:implementation}

We implemented the \tbuddy{} proof-generating BDD package by
modifying the widely used \buddy{} BDD package, developed by J{\o}rn
Lind-Nielsen in the 1990s~\cite{bryant:fmcad:2022}.  This involved
adding several additional fields to the BDD node and cache entry data
structures, yielding a total memory overhead of $1.35\times$.
\Tbuddy{} generates proofs in the LRAT proof format~\cite{cruz-cade-2017}.
We then implemented \tbsat{}, a proof-generating SAT solver based on \tbuddy{}.

\Tbsat{} supports three different evaluation mechanisms:
\begin{description}
\item[Linear:] Form the conjunction of the clauses.  No quantification
  is performed.  
  This mode matches the operation described for the original version of {\sc ebddres}~\cite{ebddres}.
  When forming the conjunction of a set of terms, the
  program makes use of a first-in, first-out queue, removing two
  elements from the front of the queue, computing their conjunction,
  and placing the result at the end of the queue.  This has the effect
  of forming a binary tree of conjunctions.
\item[Bucket Elimination:] Place the BDDs representing the clauses
  into buckets according to the levels of their topmost variables.  Then
  process the buckets from lowest to highest.  While a bucket has more
  than one element, repeatedly remove two elements, form their
  conjunction, and place the result in the bucket designated by its
  topmost variable.  Once the bucket has a single element,
  existentially quantify the topmost variable and place the result in the appropriate
  bucket~\cite{dechter-ai-1999}.  This matches the operation described for the revised version of {\sc ebddres}~\cite{Jussila:2006}.
\item[Scheduled:] Perform operations as specified by a {\em scheduling} file, as described below.
\end{description}  
The scheduling file contains a sequence of lines, each providing a
command in a simple, stack-based notation:
\begin{center}
\begin{tabular}{ll}
\texttt{c} $c_1, \ldots, c_k$ & Push the BDD representations of the specified clauses onto the stack\\
\texttt{a} $m$ &                Replace the top $m$ elements on the stack with their conjunction\\
\texttt{q} $v_1, \ldots, v_k$ & Replace the top stack element with its quantification by the specified variables\\
\end{tabular}  
\end{center}

\section{Experimental Results}
\label{sect:experimental}

In our preliminary experiments, we found that the capabillities of \tbsat{} 
differ greatly from the more mainstream CDCL
solvers, and it therefore must be evaluated by a different set of
standards.  In particular, CDCL solvers are most commonly evaluated according 
to their performance  on collections of benchmark problems in a series of annual solver
competitions.  Over the years, the benchmark problems have been
updated to provide new challenges and to better distinguish the
performance of the different solvers.  This competition has stimulated
major improvements in the solvers through improved algorithms and
implementation techniques.  One unintended consequence,
however, has been that the benchmarks have evolved to be only at, or
slightly beyond, the capabilities of CDCL solvers.

As an example, as is discused in Section~\ref{sect:urquhart},
Simon~\cite{chatalic-cade-2000} and Li~\cite{li-dam-2003} contributed
multiple benchmark formulas for the 2002 SAT
competition~\cite{simon:amai:2004} based on a class of unsatisfiable
formulas devised by Urquhart~\cite{Urquhart}.  The formulas scale
quadratically by a size parameter $m$, both in terms of the number of
variables and the number of clauses.  Simon's largest benchmark had
$m=5$, while Li's had $m=4$.  No solver at the time could complete for
these formulas, even though Li's formula for $m=4$ has only 288
variables and 768 clauses.  The 2022 SAT competition featured a
special ``Anniversary track'' using as formulas the 5355 formulas that
have been used across all prior SAT competitions.  In all, 32 solvers
participated in the competition with a 5000-second time limit for each
problem.  Even after years of improvements in the solvers and with
vastly better hardware, none of the solvers completed these
20-year-old benchmark problems.  There has been no attempt to evaluate
solvers running on Urquhart formulas for larger values of $m$, because
these were clearly beyond the reach of the competing solvers.

By
contrast,  \tbsat{} can easily handle the Urquhart
formulas.  Generating proofs of unsatisfiability for Simon's benchmark
with $m=5$ and Li's benchmark with $m=4$ requires 0.23 and 0.13
seconds, respectively.
We show
experimental results with $m=38$ for Li's version and $m=60$ for
Simon's. In a more recent
effort~\cite{bryant:fmcad:2022}, we augmented \tbsat{} to use Gaussian
elimination for reasoning about parity constraints, allowing us to
generate an unsatisfiability proof for Li's version with $m=316$, a
formula with over two million variables and five million clauses.
This example demonstrates that
measuring performance on  benchmarks designed to evaluate CDCL solvers
cannot capture the full capabilities of a BDD-based SAT solver.

In the following experiments, we explore the capability of \tbsat{} on
four scalable benchmark problems that pose major challenges for CDCL
solvers.  We do so not to show that \tbsat{} is uniformly superior,
but rather that it performs very well on some classes of problems for which CDCL is especially weak.
A long-term research direction
is to combine the capabilities of CDCL and BDDs to build on the strengths of each.

All experiments were performed on a 3.2~GHz Apple M1~Max processor
with 64~GB of memory and running the OS~X operating system.  The
runtime for each experiment was limited to 1000 seconds.
We compare the performance of \tbsat{} to that of \kissat{}, the
winner of several recent SAT solver
competitions~\cite{biere-kissat-2020}.  \Kissat{} represents the state
of the art in CDCL solvers.
The proofs
were checked using \drattrim{} for the proofs generated by \kissat{}
and \lratcheck{} for those generated by \tbsat{}.
We report both the elapsed time
by the solver, as well as the total number of clauses in the proof of
unsatisfiability.  For \kissat{}, the proof clauses indicate the conflicts the solver encountered during its search.
For \tbsat{}, these are
the defining clauses
for the extension variables (up to four per BDD node generated) and
the derived clauses (one per input clause and up to two per result
inserted into the operation cache.)

\subsection{Reordered Parity Formulas}

\begin{figure}
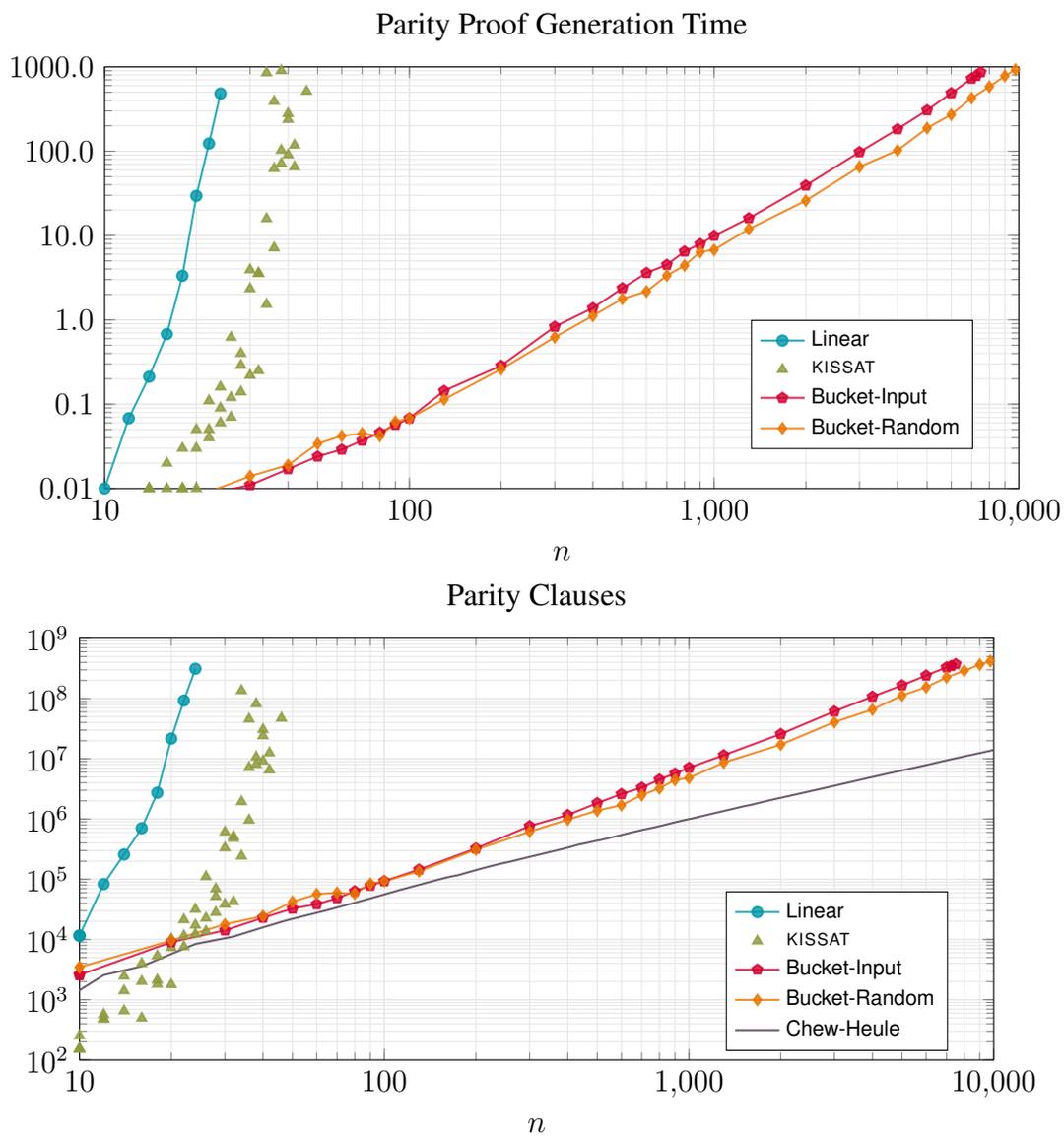

\centering{
\begin{tikzpicture}[scale = 1.0]
          \begin{axis}[mark options={scale=1},grid=both, grid style={black!10}, xmode=log, ymode=log, legend style={at={(0.95,0.40)}}, legend cell align={left},
              x post scale=1.8, xlabel=$n$,
              xmin=10,xmax=10000,ymin=0.01,ymax=1000,
              xtick={10,100,1000,10000}, xticklabels={$10$,$100$, $1{,}000$, $10{,}000$}, 
              ytick={0.01,0.1,1.0,10.0,100.0,1000.0}, yticklabels={$0{.}01$, $0{.}1$, $1{.}0$, $10{.}0$,$100{.}0$, $1000{.}0$}, 
              title={Parity Proof Generation Time}]

            \input{data-formatted/chew-linear-seconds.tex}
            \input{data-formatted/chew-kissat-seconds.tex}
            \input{data-formatted/chew-inputorder-seconds.tex}
            \input{data-formatted/chew-randomorder-seconds.tex}            

            \legend{\scriptsize \textsf{Linear}, \scriptsize \textsf{\kissat{}}, \scriptsize \textsf{Bucket-Input}, \scriptsize \textsf{Bucket-Random}}
          \end{axis}
\end{tikzpicture}

\begin{tikzpicture}[scale = 1.0]
          \begin{axis}[mark options={scale=1},grid=both, grid style={black!10}, xmode=log, ymode=log, legend style={at={(0.95,0.40)}}, legend cell align={left},
              x post scale=1.8, xlabel=$n$,
              xmin=10,xmax=10000,ymin=100,ymax=1e9,
              xtick={10,100,1000,10000}, xticklabels={$10$,$100$, $1{,}000$, $10{,}000$}, 
              title={Parity Clauses}]

            \input{data-formatted/chew-linear-clauses.tex}
            \input{data-formatted/chew-kissat-clauses.tex}
            \input{data-formatted/chew-inputorder-clauses.tex}
            \input{data-formatted/chew-randomorder-clauses.tex}            
            \input{data-formatted/chew-drat-clauses.tex}            

            \legend{\scriptsize \textsf{Linear}, \scriptsize \textsf{\kissat{}}, \scriptsize \textsf{Bucket-Input}, \scriptsize \textsf{Bucket-Random}, \scriptsize \textsf{Chew-Heule}}
          \end{axis}
\end{tikzpicture}
} 
\caption{Generating unsatisfiability proofs for reordered parity encodings with $n$ data variables}
\label{fig:data:parity}
\end{figure}  


Chew and Heule~\cite{Chew} introduced a benchmark problem based on computing
the parity of a set of Boolean values $x_1, \ldots, x_n$ using two
different orderings of the inputs, and with one of the variables
negated in the second computation:
\begin{eqnarray*}
 {\it ParityA}(x_1, \ldots, x_n) & = & x_1 \boolxor x_2 \boolxor  \cdots \boolxor x_n \\
  {\it ParityB}(x_1, \ldots, x_n) & = & \left[p_1 \boolxor x_{\pi(1)}\right] \boolxor \left[p_2 \boolxor x_{\pi(2)}\right]  \boolxor \cdots \boolxor \left[p_n \boolxor x_{\pi(n)}\right]
\end{eqnarray*}
where $\pi$ is a random permutation, and and each $p_i$ is either 0 or
1, with the restriction that $p_i = 1$ for only one value of $i$.  The two sums associate from left to right.
The
formula ${\it ParityA} \booland {\it ParityB}$ is therefore unsatisfiable,
but the permutation makes this difficult for CDCL solvers to determine.
The CNF has a total of $3n - 2$
variables---$n$ values of $x_i$, plus the auxilliary variables encoding
the intermediate terms in the two expressions.

Chew and Heule experimented with the CDCL solver {\sc CaDiCaL}~\cite{biere-cadical-2019}
and found it could not handle cases with $n$ greater than 50.
They devised a specialized method for directly generating proofs in
the DRAT proof system, obtaining proofs that scale as $O(n \log n)$
and gave results for up to $n=4{,}000$.  They also tried {\sc ebddres}, but only in its default mode, where it performs only linear evaluation without any quantification.

Figure \ref{fig:data:parity} shows the result of applying both
\tbsat{} and \kissat{} to this problem.  In this and other figures,
the top graph shows how the runtime scales with the problem size,
while the bottom graph shows how the number of proof clauses scale.
Both graphs are log-log plots,
and so the values are highly compressed along both dimensions.  
Linear evaluation performs poorly, only handling up to $n=24$ within the 1000-second time limit,
generating a proof with over 312~million clauses.  Using
\kissat{}, we found the results were very sensitive to the choice of
random permutation, and so we show results using three different random
seeds for each value of $n$.  We were able to generate proofs for
instances with $n$ up to 46 within the time limit but also started having timeouts with $n=42$.
We can see that \kissat{} does better than linear evalution with \tbsat{},
but both appear to scale exponentially.

Bucket elimination, on the other hand, displays much better scaling,
as is shown in the log-log plot of Figure \ref{fig:data:parity}. We found the best
performance was achieved by randomly permuting the variables, although
this strategy only yields a constant factor improvement over the
ordering from the CNF file.  As the graphs show, we were able to
handle cases with $n$ up to $9{,}750$, within the time limit.  This
generated a proof with over 419~million clauses, but the LRAT checker
was able to verify this proof in 256 seconds.  Although \tbsat{} could
generate proofs for larger values of $n$, these exceeded the capacity
of the LRAT checker.

Included in the second graph are results for running Chew and Heule's
proof generator on this problem.  As can be seen, the proof sizes generated
by \tbsat{} are comparable to theirs up to around $n=100$.  From there
on, however, the benefit of their $O(n\,\log n)$ algorithm becomes
apparent.  Even for $n = 10{,}000$, their proof contains less than
11~million clauses.  Of course, their construction relies on
particular properties of the underlying problem, while ours was
generated by a general-purpose SAT solver.

\subsection{Urquhart Formulas}
\label{sect:urquhart}

Urquhart~\cite{Urquhart} introduced a family of formulas that require
resolution proofs of exponential size.
Over the years, two families of
SAT benchmarks have been labeled as ``Urquhart
Problems:'' one developed by Simon~\cite{chatalic-cade-2000}, and the other by Li~\cite{li-dam-2003}.  These are considered to be difficult
challenge problems for SAT solvers.  
Here we define
their general form, describe the differences between the two families, and evaluate the
performance of both \kissat{} and \tbsat{} on both classes.  

Urquhart's construction is based on a class of bipartite graphs with
special properties.  Define $G_k$ as the set of undirected
graphs with each graph satisfying the following properties:
\begin{itemize}
\item It is {\em bipartite}:  The set of vertices can be partitioned into sets
  $L$ and $R$, such that the edges $E$ satisfy $E \subseteq L \times R$.
\item It is {\em balanced}: $|L| = |R|$.
\item It has bounded degree:  No vertex has more than $k$ incident edges.
\end{itemize}  
Furthermore, the graphs must be {\em expanders}, defined as
follows~\cite{hoory-ams-2006}.  For a subset of vertices $U
\subseteq L$, define $R(U)$ to be those vertices in $R$ adjacent to
the vertices on $U$.  A graph in $G_k$ is an expander if there
is some constant $d > 0$, such that for any $U \subset L$ with $|U|
\leq |L|/2$, the set $R(U)$ satisfies $|R(U)| \geq (d+1) |U|$.
Urquhart considers expander graphs with degree bound $k=5$ and that
are parameterized by a size value $m$ with $|L|=|R|=m^2$.

To transform such a graph into a formula,
each edge $(i, j) \in E$ has an associated variable
$x_{\{i,j\}}$.  (We use this notation to emphasize that the order of the indices does not matter.)
Each vertex is assigned a polarity $p_i \in {0,1}$,
such that the sum of the polarities is odd.
The clauses then encode the formula:
\begin{eqnarray*}
 \sum_{i=1}^{2m^2} \left[ \sum_{(i,j) \in E} x_{\{i,j\}} + p_i\right] & \equiv & 0 \pmod{2}\\
\end{eqnarray*}  
This is false, of course, since each edge gets counted twice in the sum,
and the sum of the polarities is odd.

The two families of benchmarks differ in how the graphs are
constructed.  Li's benchmarks are based on the explicit construction
of expander graphs due to
Margulis~\cite{gabber-jcss-1981,margulis-1973} that is cited by
Urquhart.  Thus, his graphs are fully defined by the size parameter
$m$.  Simon's benchmarks are based on randomly generated graphs, and
thus they are characterized by both the size parameter $m$ and the
initial random seed $s$.  Although random graphs satisfy the expander
condition with high probability~\cite{hoory-ams-2006}, it is unlikely
that the particular instances generated by Simon's benchmark
generator are truly expander graphs.  The widely used SAT benchmarks
with names of the form \verb@Urq@$M$\verb@_@$S$\verb@.cnf@ were
generated by Simon's program for size parameter $m=M$ and initial seed
$S$.  For Simon's benchmarks, we  used five different seeds for each value of $m$.

\begin{figure}
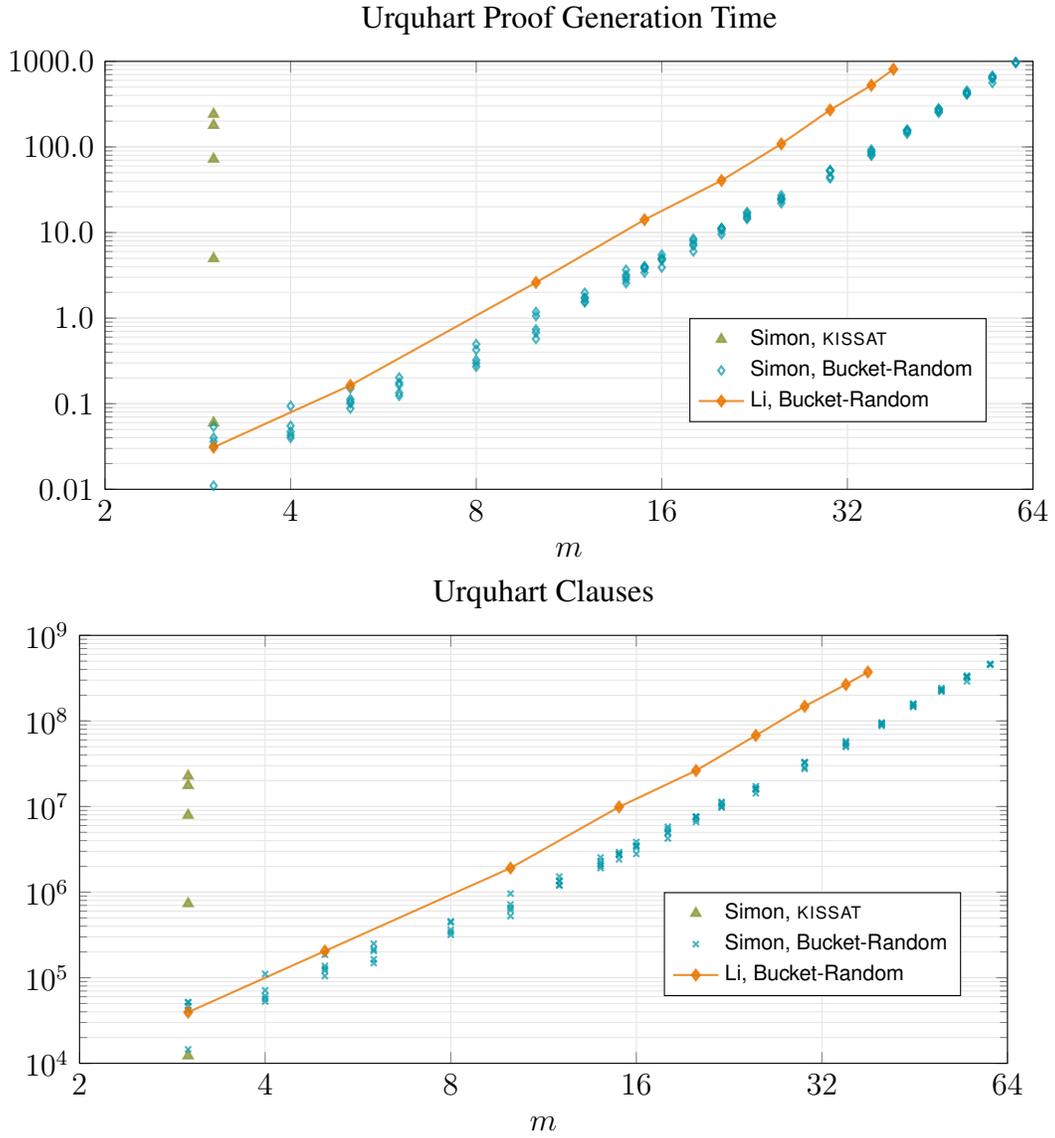

\centering{
\begin{tikzpicture}[scale = 1.0]
          \begin{axis}[mark options={scale=1},grid=both, grid style={black!10}, xmode=log, ymode=log, legend style={at={(0.95,0.40)}}, legend cell align={left},
              x post scale=1.8, xlabel=$m$,
                              xtick={2,4,8,16,32,64}, xticklabels={$2$, $4$,$8$,$16$,$32$,$64$},
                              xmin=2,xmax=64,ymin=0.01,ymax=1000,
              ytick={0.01,0.1,1.0,10.0,100.0,1000.0}, yticklabels={$0{.}01$, $0{.}1$, $1{.}0$, $10{.}0$,$100{.}0$, $1000{.}0$}, 
                              title={Urquhart Proof Generation Time}]
            \input{data-formatted/urquhart-simon-kissat-seconds.tex}
            \input{data-formatted/urquhart-simon-randomorder-seconds.tex}            
            \input{data-formatted/urquhart-li-randomorder-seconds.tex}            
            \legend{ \scriptsize \textsf{Simon, \kissat},
                     \scriptsize \textsf{Simon, Bucket-Random},
                     \scriptsize \textsf{Li, Bucket-Random},
            }
          \end{axis}
\end{tikzpicture}

\begin{tikzpicture}[scale = 1.0]
          \begin{axis}[mark options={scale=1},grid=both, grid style={black!10}, xmode=log, ymode=log, legend style={at={(0.95,0.40)}}, legend cell align={left},
              x post scale=1.8, xlabel=$m$,
                              xtick={2,4,8,16,32,64}, xticklabels={$2$, $4$,$8$,$16$,$32$,$64$},
                              xmin=2,xmax=64,ymin=10000,ymax=1e9, title={Urquhart Clauses}]

            \input{data-formatted/urquhart-simon-kissat-clauses.tex}
            \input{data-formatted/urquhart-simon-randomorder-clauses.tex}            
            \input{data-formatted/urquhart-li-randomorder-clauses.tex}            
            \legend{ \scriptsize \textsf{Simon, \kissat},
              \scriptsize \textsf{Simon, Bucket-Random},
              \scriptsize \textsf{Li, Bucket-Random},
            }
          \end{axis}
\end{tikzpicture}
} 
\caption{Generating unsatisfiability proofs for Urquhart formulas with size parameter $m$.  \Kissat{} timed out for even the minimum-sized version of Li's benchmark ($m=3$).}
\label{fig:data:urquhart}
\end{figure}  

Figure \ref{fig:data:urquhart} shows data for running \kissat{}, as well as \tbsat{} using bucket elimination.
The data for \kissat{} demonstrates how
difficult these benchmark problems are for CDCL solvers.  With a time limit of 1000 seconds,
we found that \kissat{} could
handle all five instances of Simon's benchmarks with $m = 3$, 
but none for larger values of $m$.  For Li's benchmarks
it failed for even the minimum case of $m=3$.
Running \tbsat{} with bucket elimination with a random ordering of the variables fares much better.
For Li's benchmarks,
it successfully handled instances up to $m=38$, yielding a proof with around 373 million clauses.
For Simon's
benchmarks, bucket elimination handled benchmarks for all five seeds up to $m=60$.
We can also see that Simon's benchmarks are decidedly easier than Li's, requiring up to
an order of magnitude fewer clauses in the proofs.

Jussila, Sinz, and Biere~\cite{Jussila:2006} showed benchmark results
for what appear to be Simon's Urquhart formulas up to $m=8$ with
performance (in terms of proof size) comparable to ours.  Indeed, in using bucket elimination,
we are replicating their approach.  We know of no prior
proof-generating SAT solver that could handle Urquhart formulas of this
scale.

\subsection{Mutilated Chessboard}

\begin{figure}
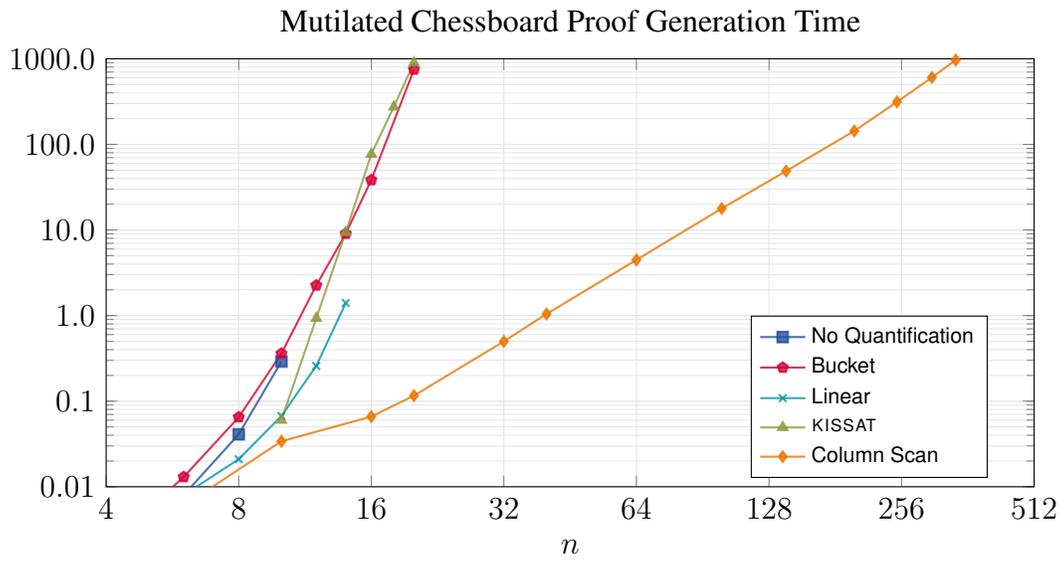
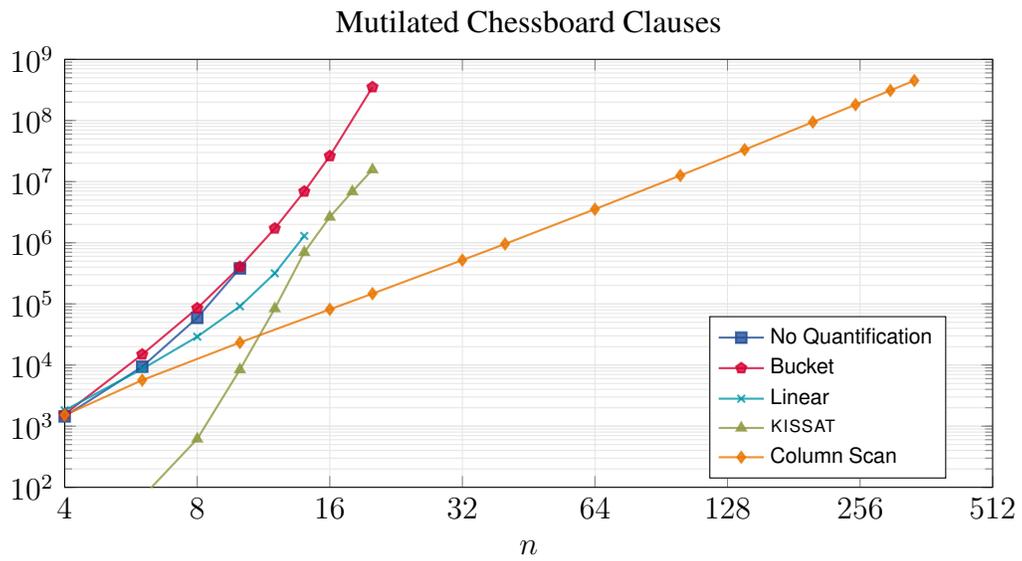

\centering{
\begin{tikzpicture}[scale = 1.0]
          \begin{axis}[mark options={scale=1.0},grid=both, grid style={black!10}, xmode=log, ymode=log, legend style={at={(0.95,0.40)}}, legend cell align={left},
                              x post scale=1.8, xlabel=$n$, xtick={4,8,16,32,64,128,256,512}, xticklabels={$4$,$8$,$16$,$32$,$64$,$128$,$256$,$512$},
                              xmin=4,xmax=512,ymin=0.01,ymax=1000, title={Mutilated Chessboard Proof Generation Time},
                              ytick={0.01,0.1,1.0,10.0,100.0,1000.0}, yticklabels={$0{.}01$, $0{.}1$, $1{.}0$, $10{.}0$,$100{.}0$, $1000{.}0$}, 
            ]
            \input{data-formatted/chess-noquant-seconds}
            \input{data-formatted/chess-bucket-seconds}
            \input{data-formatted/chess-linear-seconds}
            \input{data-formatted/chess-kissat-seconds}
            \input{data-formatted/chess-scan-seconds}
            
            \legend{\scriptsize \textsf{No Quantification}, \scriptsize \textsf{Bucket}, \scriptsize \textsf{Linear}, \scriptsize \textsf{\textsc{kissat}}, \scriptsize \textsf{Column Scan}}
          \end{axis}
\end{tikzpicture}

\begin{tikzpicture}[scale = 1.0]
          \begin{axis}[mark options={scale=1.0},grid=both, grid style={black!10}, xmode=log, ymode=log, legend style={at={(0.95,0.40)}}, legend cell align={left},
                              x post scale=1.8, xlabel=$n$, xtick={4,8,16,32,64,128,256,512}, xticklabels={$4$,$8$,$16$,$32$,$64$,$128$,$256$,$512$},
                              xmin=4,xmax=512,ymin=100,ymax=1e9, title={Mutilated Chessboard Clauses}]
            \input{data-formatted/chess-noquant-clauses}
            \input{data-formatted/chess-bucket-clauses}
            \input{data-formatted/chess-linear-clauses}
            \input{data-formatted/chess-kissat-clauses}
            \input{data-formatted/chess-scan-clauses}
            
            \legend{\scriptsize \textsf{No Quantification}, \scriptsize \textsf{Bucket}, \scriptsize \textsf{Linear}, \scriptsize \textsf{\textsc{kissat}}, \scriptsize \textsf{Column Scan}}
          \end{axis}
\end{tikzpicture}
} 
\caption{Generating unsatisfiability proofs for $n \times n$ mutilated  chess boards}
\label{fig:data:chess}
\end{figure}  


The mutilated chessboard problem considers an $n \times n$ chessboard,
with the corners on the upper left and the lower right removed.  It
attempts to tile the board with dominos, with each domino covering two
squares.  Since the two removed squares had the same color, and each
domino covers one white and one black square, no tiling is possible.
This problem has been well studied in the context of resolution
proofs, for which it can be shown that any proof must be of
exponential size~\cite{Alekhnovich}.

A standard CNF encoding involves defining Boolean variables to
represent the boundaries between adjacent squares, set to 1 when a
domino spans the two squares, and set to 0 otherwise.  The clauses
then encode an Exactly1 constraint for each square, requiring each
square to share a domino with exactly one of its neighbors.  We
label the variables representing a horizontal boundary
between a square and the one below as $y_{i,j}$, with $1 \leq i < n$ and $1 \leq j
\leq n$.  The variables representing the vertical boundaries
are labeled $x_{i,j}$, with $1 \leq i \leq n$ and
$1 \leq j< n$.  With a mutilated chessboard, we have 
$y_{1,1} = x_{1,1} = y_{n-1,n} = x_{n,n-1} = 0$.  

As the plots of Figure \ref{fig:data:chess} show, a straightforward
application of linear conjunctions or bucket elimination by \tbsat{}
displays exponential scaling.  Indeed, \tbsat{} fares no better than
\kissat{} when operating in either of these modes, with all limited to
$n \leq 20$ within the 1000-second time limit.

On the other hand, another approach, inspired by symbolic model
checking~\cite{burch-ic-1992}, demonstrates far better scaling, reaching $n=340$.  It is
based on the following observation: when processing the columns from
left to right, the only information required to place dominos in
column $j$ is the identity of those rows $i$ for which a domino
crosses horizontally from $j-1$ to $j$.  This information is encoded
in the values of $x_{i,j-1}$ for $1 \leq i \leq n$.

In particular, group the variables into columns, with $X_j$ denoting
variables $x_{1,j}, \ldots, x_{n,j}$, and $Y_j$ denoting variables
$y_{1,j}, \ldots, y_{n-1,j}$.  Scanning the board from left to right,
consider $X_j$ to encode the ``state'' of processing after completing
column $j$.
As the scanning process reaches column $j$,
there is a {\em characteristic function} $\sigma_{j-1}(X_{j-1})$
  describing the set of allowed crossings of horizontally-oriented
  dominos from column $j-1$ into column $j$.  No other information
  about the configuration of the board to the left is required.  The
  characteristic function after column $j$ can then be computed as:
  \begin{eqnarray}
    \sigma_{j}(X_j)  & = & \exists X_{j-1}
    \; \bigl[ \sigma_{j-1}(X_{j-1}) \booland \exists Y_{j} \;T_j(X_{j-1}, Y_{j}, X_{j}) \bigr] \label{eqn:chess:transition}
  \end{eqnarray}
where $T_j(X_{j-1}, Y_{j}, X_{j})$ is a ``transition relation''
consisting of the conjunction of the Exactly1 constraints for column
$j$.  From this, we can existentially quantify the variables $Y_{j}$
to obtain a BDD encoding all compatible combinations of the variables
$X_{j-1}$ and $X_{j}$.  By conjuncting this with the characteristic
function for column $j-1$ and existentially quantifying the variables
$X_{j-1}$, we obtain the characteristic function for column $j$.  With
a mutilated chessboard, we generate leaf node $L_0$ in attempting the
final conjunction.  Note that Equation~\ref{eqn:chess:transition} does
not represent a reformulation of the mutilated chessboard problem.
It simply defines a way to schedule the conjunction and
quantification operations over the input clauses.

One important rule-of-thumb in symbolic model checking is that the
successive values of the next-state variables must be adjacent in the
variable ordering.  Furthermore, the vertical variables in $Y_j$ must
be close to their counterparts in $X_{j-1}$ and $X_{j}$.  Both
objectives can be achieved by ordering the variables row-wise,
interleaving the variables $x_{i,j}$ and $y_{i,j}$, ordering first by
row index $i$ and then by column index $j$.  This requires the
quantification operations of Equation~\ref{eqn:chess:transition} to be
performed on non-root variables.

In our experiments, we found that this scanning reaches a fixed point
after processing $n/2$ columns.  That is, from that column onward, the
characteristic functions become identical, except for a renaming of
variables.  This indicates that the set of all possible horizontal
configurations stabilizes halfway across the board.  Moreover, the BDD
representations of the states grow as $O(n^2)$.  For $n=340$
largest has just $29{,}239$ nodes.
The problem size for the mutilated chessboard scales as $n^2$, 
the number of squares in the board.  Thus, an instance with $n=340$
is  $289$ times larger than an instance with $n=20$, in terms of the
number of input variables and clauses.  Column scanning yields a major benefit in the solver performance.

The plot 
labeled ``No Quantification'' demonstrates the importance of
including existential quantification in solving this problem.  These
data were generated by using the same schedule as with column
scanning, but with all quantification operations omitted.  As can be
seen, this approach could not scale beyond $n=10$.

It is interesting to reflect on how our column-scanning approach
relates to SAT-based bounded model checking
(BMC)~\cite{biere:tacas:1999}.  This approach to verification encodes 
the operation of a state transition system for $k$ steps, for some fixed value of $k$,
by
replicating the transition relation $k$
times.  It then uses a SAT solver to
detect whether some condition can arise within $k$ steps of operation.
By contrast, we effectively
compress the mutilated chessboard problem into a state machine that
adds tiles to successive columns of the board and then perform a
BDD-based reachability computation for this system, much as would a symbolic model checker~\cite{burch-ic-1992}.
Just as BDD-based
model checking can outperform SAT-based BMC for some problems, we have
demonstrated that a BDD-based SAT solver can sometimes outperform a
search-based SAT solver.

\subsection{Pigeonhole Problem}

\begin{figure}
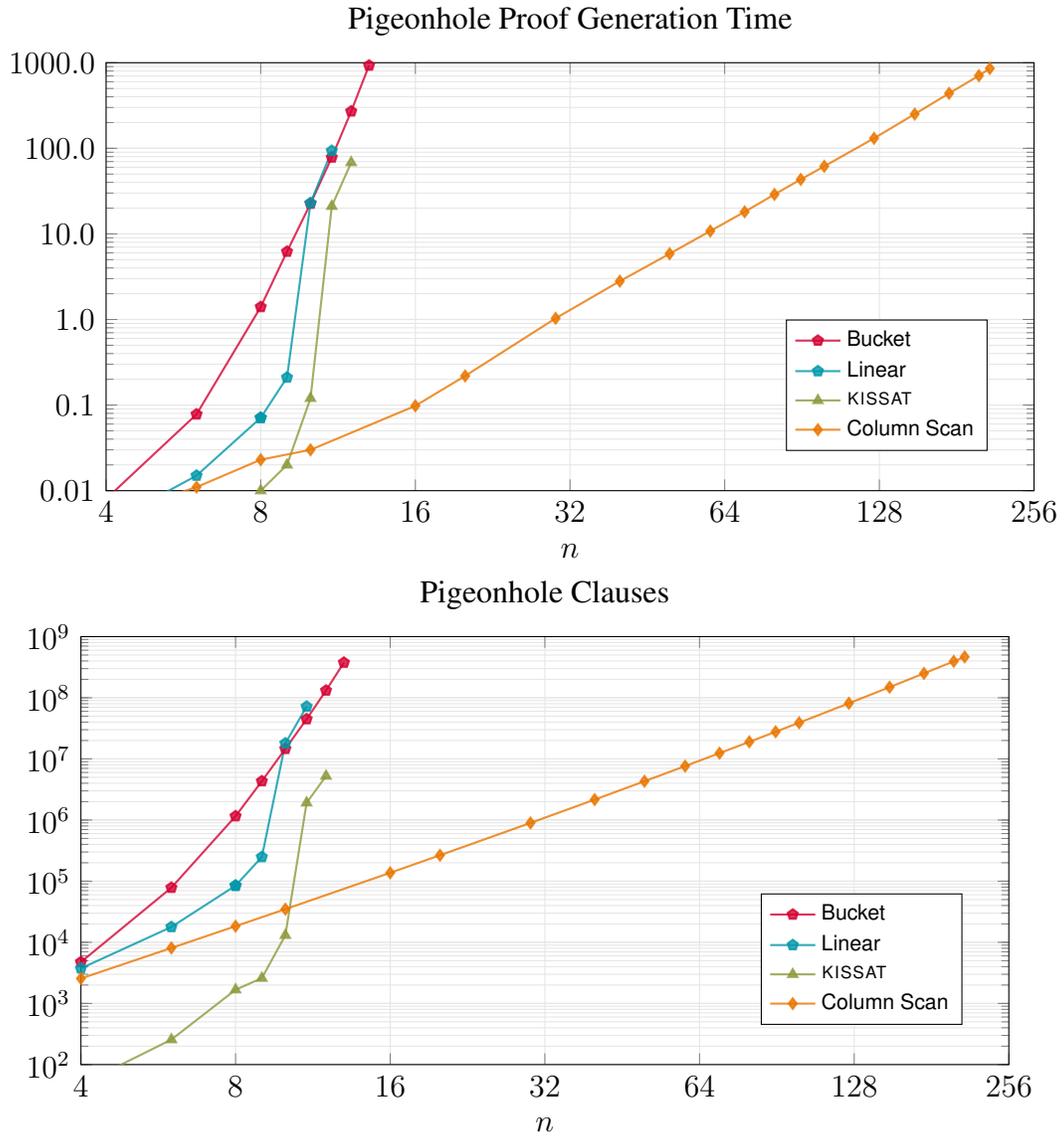

\centering{
\begin{tikzpicture}[scale = 1.0]
          \begin{axis}[mark options={scale=1.0},grid=both, grid style={black!10}, xmode=log, ymode=log, legend style={at={(0.95,0.40)}}, legend cell align={left},
                              x post scale=1.8, xlabel=$n$, xtick={4,8,16,32,64,128,256,512}, xticklabels={$4$,$8$,$16$,$32$,$64$,$128$,$256$,$512$},
                              xmin=4,xmax=256,ymin=0.01,ymax=1000, title={Pigeonhole Proof Generation Time},
                              ytick={0.01,0.1,1.0,10.0,100.0,1000.0}, yticklabels={$0{.}01$, $0{.}1$, $1{.}0$, $10{.}0$,$100{.}0$, $1000{.}0$}, 
            ]
            \input{data-formatted/pigeon-sinz-bucket-seconds}
            \input{data-formatted/pigeon-sinz-linear-seconds}
            \input{data-formatted/pigeon-sinz-kissat-seconds}
            \input{data-formatted/pigeon-sinz-scan-seconds}
            
            \legend{\scriptsize \textsf{Bucket}, \scriptsize \textsf{Linear}, \scriptsize \textsf{\textsc{kissat}}, \scriptsize \textsf{Column Scan}}
          \end{axis}
\end{tikzpicture}

\begin{tikzpicture}[scale = 1.0]
          \begin{axis}[mark options={scale=1.0},grid=both, grid style={black!10}, xmode=log, ymode=log, legend style={at={(0.95,0.40)}}, legend cell align={left},
                              x post scale=1.8, xlabel=$n$, xtick={4,8,16,32,64,128,256,512}, xticklabels={$4$,$8$,$16$,$32$,$64$,$128$,$256$,$512$},
                              xmin=4,xmax=256,ymin=100,ymax=1e9, title={Pigeonhole Clauses}]
            \input{data-formatted/pigeon-sinz-bucket-clauses}
            \input{data-formatted/pigeon-sinz-linear-clauses}
            \input{data-formatted/pigeon-sinz-kissat-clauses}
            \input{data-formatted/pigeon-sinz-scan-clauses}
            
            \legend{\scriptsize \textsf{Bucket}, \scriptsize \textsf{Linear}, \scriptsize \textsf{\textsc{kissat}}, \scriptsize \textsf{Column Scan}}
          \end{axis}
\end{tikzpicture}
} 
\caption{Generating unsatisfiability proofs for assigning $n+1$ pigeons to $n$ holes using Sinz's encoding}
\label{fig:data:pigeon}
\end{figure}


The pigeonhole problem is one of the most studied problems in
propositional reasoning.  Given a set of $n$ holes and a set of $n+1$
pigeons, it asks whether there is an assignment of pigeons to holes
such that 1) every pigeon is in some hole, and 2) every hole contains
at most one pigeon.  The answer is no, of course, but any resolution
proof for this must be of exponential length~\cite{Haken:1985}.  
Groote and Zantema have
shown that any BDD-based proof of the principle that only uses the
Apply algorithm must be of exponential size~\cite{groote-dam-2003}.
On the other hand, Cook
constructed an extended resolution proof of size $O(n^4)$, in part to
demonstrate the expressive power of extended resolution~\cite{Cook:1976}.

We used an representation of the problem that scales as as $O(n^2)$,
using an encoding of the at-most-one constraints due to
Sinz~\cite{card}.
It starts with a set of
variables $p_{i,j}$ for $1 \leq i \leq n$ and $1 \leq j \leq n+1$,
with the interpretation that pigeon $j$ is assigned to hole $i$.
Encoding the property that each pigeon $j$ is assigned to some hole can be
expressed as with a single clause:
\begin{eqnarray*}
{\it Pigeon}_j & = & \bigvee_{i = 1}^{n} p_{i,j}
\end{eqnarray*}

Sinz's method of encoding the property that each hole $i$ contains at most one pigeon
 introduces auxilliary variables to
effectively track which holes are occupied,
starting with pigeon 1 and working upward.
These variables are labeled $s_{i,j}$ for $1 \leq i \leq n$
and $1 \leq j \leq n$.
Informally, variables $s_{i,1}, s_{i,2}, \ldots, s_{i,n}$ serves as a 
{\em signal chain} that indicates the point at which a pigeon has been assigned to hole $i$.
For each hole $i$, there is a total of
$3n-1$ clauses:
\begin{center}
  \smallskip
  \begin{tabular}{lccl}
\toprule
    \multicolumn{1}{c}{Effect} & Formula & Clause & Range \\
\midrule
    Generate & $p_{i,j} \rightarrow s_{i,j}$ & $\obar{p}_{i,j}\,s_{i,j}$ & $1 \leq j \leq n$ \\
    Propagate & $s_{i,j-1} \rightarrow s_{i,j}$ & $\obar{s}_{i,j-1}\,s_{i,j}$ & $1 <  j \leq n $ \\
    Suppress & $s_{i,j-1} \rightarrow \obar{p}_{i,j}$ & $\obar{s}_{i,j-1} \, \obar{p}_{i,j}$ & $1 < j \leq n+1$ \\
\bottomrule
\end{tabular}
  \smallskip
\end{center}
Each of these clauses serves either to define how the next value in the chain
is to be computed, or to describe the effect of the signal on the
allowed assignments of pigeons to the hole.  That is, for hole $i$,
the signal is {\em generated} at position $j$ if pigeon $j$ is assigned
to that hole.  Once set, the signal continues to {\em propagate} across
higher values of $j$.  Once the signal is set, it {\em suppresses}
further assignments of pigeons to the hole.
This encoding require $3n-1$ clauses and $n$ auxilliary variables per hole.

Figure~\ref{fig:data:pigeon} shows the results of running the two
solvers on this problem.  Once again we see \tbsat{} with either
linear or bucket evaluation having exponential scaling, as does
\kissat{}.  None can go beyond $n=13$ within the 1000-second time
limit.

On the other hand, the
column scanning approach used for the mutilated checkerboard can also
be applied to the pigeonhole problem when the Sinz encoding is
used.  Consider an array with hole $i$ represented by row $i$ and
pigeon $j$ represented by column $j$.  Let $S_{j}$
represent the auxilliary variables $s_{i,j}$
for $1 \leq i \leq n$.  The ``state'' is then encoded in these auxilliary
variables.
In processing pigeon $j$, we can assume that the possible
combinations of values of auxilliary variables $S_{j-1}$  is
encoded by a characteristic function $\sigma_{j-1}(S_{j-1})$.
In addition, we incorporate into this characteristic function the
requirement that each pigeon $k$, for $1 \leq k \leq j-1$ is assigned
to some hole.  Letting $P_j$ denote the variables $p_{i,j}$ for $1
\leq i \leq n$, the characteristic function at column $j$ can then be
expressed as:
  \begin{eqnarray}
    \sigma_{j}(S_j)  & = & \exists S_{j-1} \; 
    \bigl[ \sigma_{j-1}(S_{j-1}) \booland \exists P_{j} \;T_j(S_{j-1}, P_{j}, S_{j}) \bigr] \label{eqn:pigeonhole:transition}
  \end{eqnarray}
where the ``transition relation'' $T_j$ consists of the clauses
associated with the auxilliary variables, plus the clause encoding
constraint ${\it Pigeon}_j$.  As with the mutilated chessboard, having
a proper variable ordering is critical to the success of a column
scanning approach.  We interleave the ordering of the variables
$p_{i,j}$ and $s_{i,j}$, ordering them first by $i$ (holes) and then
by $j$ (pigeons.)

Figure \ref{fig:data:pigeon} demonstrates the effectiveness of the
column-scanning approach.  We were able to handle instances up to
$n=210$.  Unlike with the mutilated chessboard, the scanning does not
reach a fixed point.  Instead, the BDDs start very small, because they
must encode the locations of only a small number of occupied holes.
They reach their maximum size at pigeon $n/2$, as the number of
combinations for occupied and unoccupied holes reaches its maximum of
$C(n, n/2)$.  The BDD sizes then drop off, symmetrically to the first
$n/2$ pigeons, as the encoding needs to track the positions of a
decreasing number of unoccupied holes.  Fortunately, all of these BDDs
scale quadraftically with $n$, reaching a maximum of $11{,}130$ nodes
for $n = 210$.

We also ran experiments using a {\em direct} encoding of the
at-most-one constraints, having a clause $\obar{p}_{i,j} \lor
\obar{p}_{i,k}$ for each hole $i$ and for $1 \leq j < k \leq n+1$.
This encoding scales as $\Theta(n^3)$.  With this encoding, we were
unable to find any method that avoided exponential scaling using either
\tbsat{} or \kissat{}.

\subsection{Evaluation}

Overall, our results demonstrate the potential for generating small
proofs of unsatisfiability using BDDs.  We were able to greatly
outperform traditional, CDCL solvers for four well-known challenge
problems.

The success for the first two benchmark problems relies on the
ability of BDDs to handle exclusive-or operations efficiently.
Generally, the exclusive-or of $k$ variables can be expressed as a BDD
with $2k+1$ nodes, including the leaves.  These representations are
also independent of the variable ordering.  As we saw, however, it is
critical to quantify variables whenever possible, to avoid requiring
the BDD to encode the parity relationships among many overlapping
subsets of the variables.  We found that bucket elimination works well
on these problems, and that randomness in the problem structure and
the variable ordering did not adversely affect performance.  This
strategy was outlined by Jussila, Sinz, and Biere; our experimental
results serve as a demonstration of the utility of their work.

The success of column scanning for the final two benchmark problems
relies on finding a way to scan in one dimension, encoding the
``state'' of the scan in a compact form.  This strategy only works
when the problem is encoded in a way that it can be partitioned along
two dimensions.  This approach draws its inspiration from symbolic
model checking, and it requires the more general capability to handle
quantification that we have presented.  One strength of modern SAT
solvers is that they generally succeeed without any special guidance
from the user.  It remains an open question whether column scanning
can be made more general and whether a suitable schedule and variable
ordering can be generated automatically.  Without these capabilities,
our results for column scanning show promise, but they require too
much guidance from the user.

Other studies have compared BDDs to CDCL solvers on a variety of
benchmark problems.  Several of these observed exponential performance
for BDD-based solvers for problems for which we have obtained more promising results.
Uribe and Stickel~\cite{uribe-1994} ran
experiments with the mutilated chessboard problem, but they did not do
any variable quantification.  Pan and Vardi~\cite{pan-sat-2004}
applied a variety of scheduling and variable ordering strategies for
the mutilated chessboard and pigeonhole problems.  Although they found
that they could get better performance than with a CDCL
solver, their performance still scaled exponentially.  Obtaining
scalability requires devising more problem-specific
approaches than the ones they considered.  Our experiments with \kissat{}
confirm that a BDD-based SAT solver requires careful attention to the
problem encoding, the variable ordering, and the use of quantification
in order to outperform a state-of-the CDCL solver.


\begin{table}
\caption{Summary data for the largest parity and Urquhart formulas solved}
\label{tab:statistics:a}
\begin{center}
\begin{tabular}{lrr}
\toprule
 \makebox[30mm][l]{Instance}
          & \multicolumn{1}{c@{~~~~~~~~~~~~~~~~~~}}{Parity-9750} & \multicolumn{1}{c}{Urquhart-Li-38} \\
\midrule
 Input variables              &     29,244   & 29,868 \\
 Input clauses                &     77,984   & 79,648 \\
\midrule
 Total BDD nodes              &  62,722,228  & 55,763,704 \\
\midrule
 Total clauses                &  419,255,800 & 372,999,366 \\
 Maximum live clauses         &  166,706,941 & 148,101,720 \\
\midrule
 Solver time (secs)              & 928.4        & 809.5 \\
 Checking time (secs)         & 288.3          & 258.5 \\
\bottomrule 
\end{tabular}      
\end{center}  
\end{table}

\begin{table}
\caption{Summary data for the largest chess and pigeonhole problems solved}
\label{tab:statistics:b}
\begin{center}
\begin{tabular}{lrr}
\toprule
 \makebox[30mm][l]{Instance}
          & \multicolumn{1}{c@{~~~~~~~~~~~~~~~~~~}}{Chess-340} & \multicolumn{1}{c}{Pigeon-Sinz-210} \\
\midrule
 Input variables              &     230,516 & 88,410 \\
 Input clauses                &     805,112 & 132,301 \\
\midrule
 Total BDD nodes              &  100,804,928 & 53,093,749 \\
\midrule
 Total clauses                &  449,676,065 & 465,887,970 \\
 Maximum live clauses         &  119,957,540 &  30,295,942 \\
\midrule
 Solver time (secs)              & 969.4      & 857.2 \\
 Checking time (secs)         & 298.0      & 340.4 \\
\bottomrule 
\end{tabular}      
\end{center}  
\end{table}

Tables~\ref{tab:statistics:a}--\ref{tab:statistics:b} provide some performance data for the
largest instances solved for each of the four benchmark problems.
A first observation is
that these problems are very large, with tens of thousands of input
variables and clauses.

Looking at the BDD data, the
total number of BDD nodes indicates the total number generated by the
function \opname{GetNode}, and for which extension variables are
created.  These are numbered in the millions, and far exceed the
number of input variables.

The entries for ``Maximum live clauses'' shows the peak number of
clauses that had been added but not yet deleted across the entire
proof.  As can be seen, these can vary from 7\% to nearly 40\% of the
total clauses.  The peak number of live clauses proved to be a limiting factor for the LRAT proof checker..

\begin{figure}
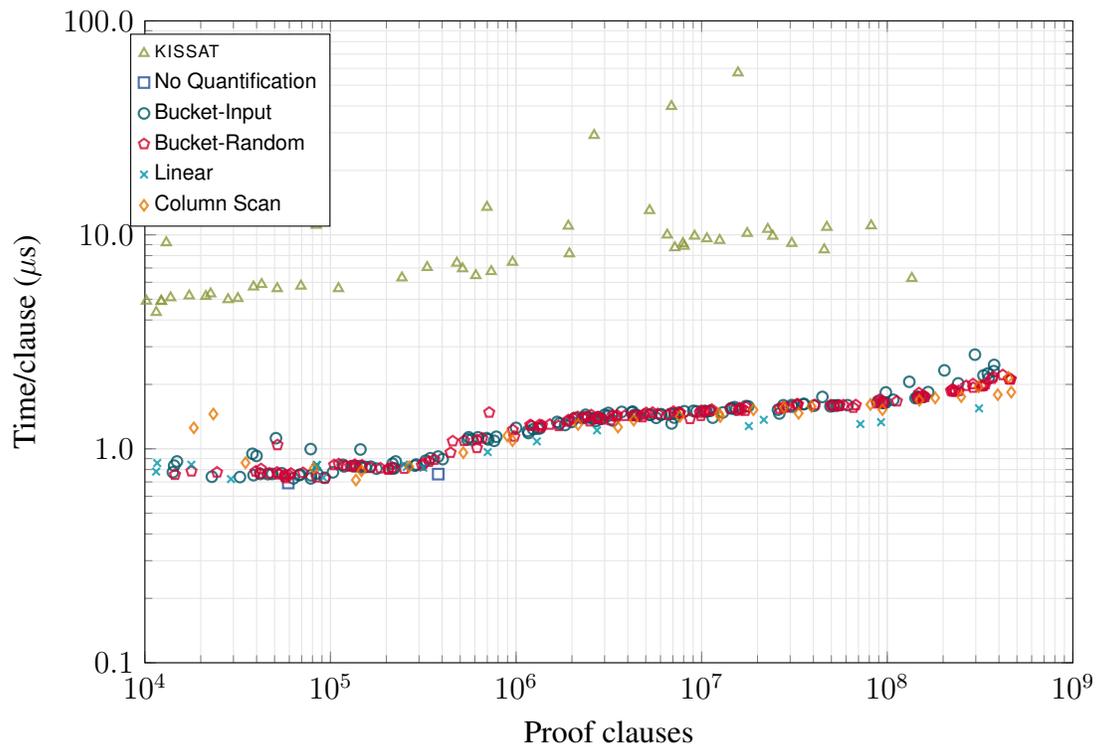

\centering{
\begin{tikzpicture}[scale = 1.0]
          \begin{axis}[mark options={scale=1.0},grid=both, grid style={black!10}, xmode=log,
              ymode=log,
              legend style={at={(0.20,0.98)}}, legend cell align={left},
                              x post scale=1.8,
                              y post scale=1.5,
                              xlabel=Proof clauses , xmin = 1e4, xmax = 1e9,
                              ymin=1e-7,ymax=1e-4,
                              ytick = {1e-7, 1e-6, 1e-5, 1e-4},
                              yticklabels = {$0.1$, $1.0$, $10.0$, $100.0$},
                              ylabel={Time/clause ($\mu$s)},
            ]
            \input{data-formatted/all-kissat-pair}
            \input{data-formatted/all-noquant-pair}
            \input{data-formatted/all-inputorder-pair}
            \input{data-formatted/all-randomorder-pair}
            \input{data-formatted/all-linear-pair}
            \input{data-formatted/all-scan-pair}
            
            \legend{
              \scriptsize \textsf{\textsc{kissat}},
              \scriptsize \textsf{No Quantification}, \scriptsize \textsf{Bucket-Input},
              \scriptsize \textsf{Bucket-Random}, \scriptsize \textsf{Linear},
              \scriptsize \textsf{Column Scan}}
          \end{axis}
\end{tikzpicture}
} 
\caption{Average generation time ($\mu$s) per clause across all benchmarks}
\label{fig:data:clause:generation}
\end{figure}  

\begin{figure}
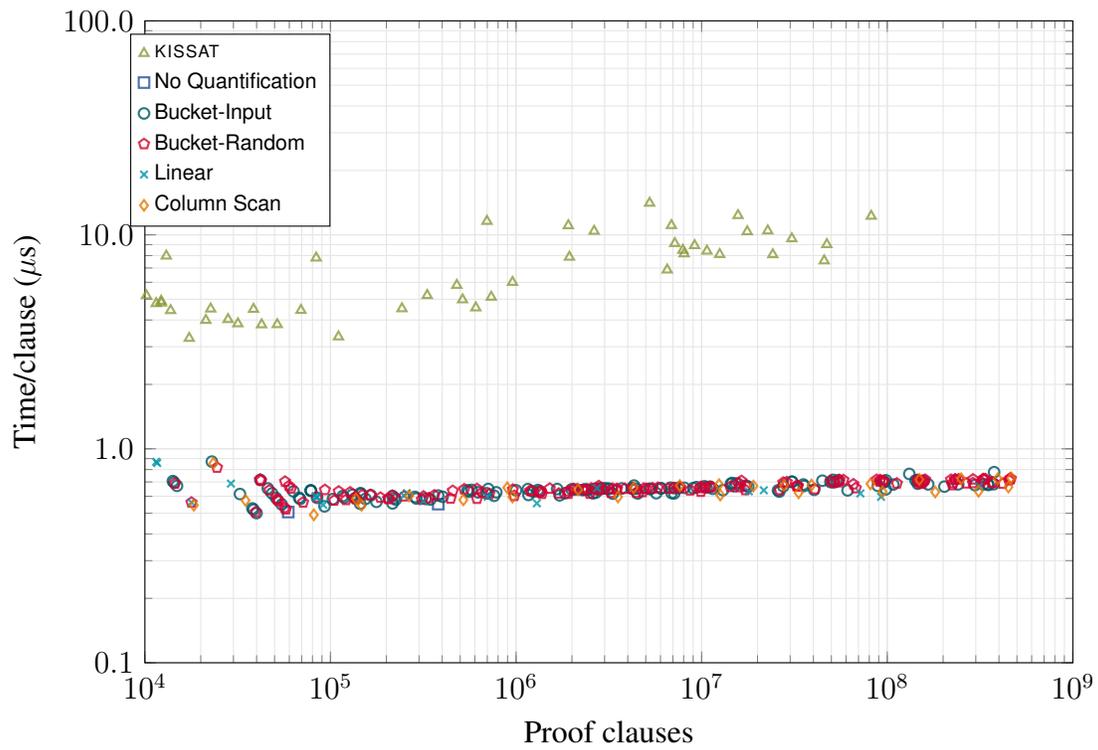

\centering{
\begin{tikzpicture}[scale = 1.0]
          \begin{axis}[mark options={scale=1.0},grid=both, grid style={black!10}, xmode=log,
              ymode=log,
              legend style={at={(0.20,0.98)}}, legend cell align={left},
                              x post scale=1.8,
                              y post scale=1.5,
                              xlabel=Proof clauses , xmin = 1e4, xmax = 1e9,
                              ymin=1e-7,ymax=1e-4,
                              ytick = {1e-7, 1e-6, 1e-5, 1e-4},
                              yticklabels = {$0.1$, $1.0$, $10.0$, $100.0$},
                              ylabel={Time/clause ($\mu$s)},
            ]
            \input{data-formatted/all-kissat-check-pair}
            \input{data-formatted/all-noquant-check-pair}
            \input{data-formatted/all-inputorder-check-pair}
            \input{data-formatted/all-randomorder-check-pair}
            \input{data-formatted/all-linear-check-pair}
            \input{data-formatted/all-scan-check-pair}
            
            \legend{
              \scriptsize \textsf{\textsc{kissat}},
              \scriptsize \textsf{No Quantification}, \scriptsize \textsf{Bucket-Input},
              \scriptsize \textsf{Bucket-Random}, \scriptsize \textsf{Linear},
              \scriptsize \textsf{Column Scan}}
          \end{axis}
\end{tikzpicture}
} 
\caption{Average checking time ($\mu$s) per clause across all benchmarks}
\label{fig:data:checking}
\end{figure}

Figure~\ref{fig:data:clause:generation} provides more insight into the
nature of the proofs generated by the CDCL solver \kissat{} and the
BDD-based solver \tbsat{}.  Each point indicates one benchmark run,
with the value on the Y axis indicating the runtime of the solver divided by the number of clauses generated, scaled
by $10^6$, while the X value is the proof size.  In other words, the Y values show the average time, in microseconds
for each proof clause to be generated.  In all, 340 points are shown,
with 75 for \kissat{}, and the rest for \tbsat{} in its various
operating modes.

These data quantify a fundamental difference between how proofs are
generated with a CDCL solver, versus with a BDD-based solver.  A CDCL
solver emits a clause each time it encounters a conflict during the
search.  This may come after many steps involving selecting a decision
variable and performing Boolean constraint propagation.  Thus, there
can be considerable, and highly variable amounts of processing between
successive clause emissions.  We see that average times ranging
between 4 and 60 microseconds for the \kissat{} runs, and even these
averages mask the considerable variations that can occur within a single run.

With a BDD-based solver, on the other hand, the proof has the form of a log
describing the recursive steps taken by the BDD algorithm, expressed
within a standard proof framework.
There is very little
variability from one run to the next, and the different
evaluation modes having minimal impact.  The only trend of note is a
general increase of the average time per clause as the proofs get
longer.  The short runs require less than 1.0 microsecond per clause,
while the longer ones require over 2.0.  This increase can be
attributed to the complexity of managing long BDD computations,
requiring garbage collection, table resizing, and other overhead
operations.

Figure~\ref{fig:data:checking} shows a similar plot, but with the Y
axis indicating the average time for the proof checkers to check each
clause.  Again, we see two important characteristics.  The proof steps
generated by \kissat{} do not include lists of antecedent clauses
(hints).  Instead, the checking program \drattrim{} scans the set of
clauses and constructs each hint sequence.  This takes significant
effort and can vary greatly across benchmarks.  The proofs generated
by \tbsat{}, on the other hand, contain full hints and can therefore
be readily checked at an average of around 0.7 $\mu$~seconds per proof
clause, regardless of the proof size or solution method.

\section{Conclusion}

The pioneering work by Biere, Sinz, and
Jussila~\cite{Jussila:2006,ebddres} did not lead to as much follow-on
work as it deserved.  Here, over fifteen years later, we found that small
modifications to their approach enable a powerful, BDD-based SAT
solver to generate proofs of unsatisfiability.  The key to its success
is the ability to perform arbitrary existential quantification.  As the
experimental results demonstrate, such a capability is critical to
obtaining reasonable performance.

More advanced BDD-based SAT solvers employ additional techniques to
improve their performance.  Extending our methods to handle these
techniques would be required to have them generate proofs of
unsatisfiability.  Some of these would be straightforward.  For
example, Weaver, Franco, and Schlipf~\cite{weaver_jsat_2006} derive a
very general set of conditions under which existential quantification
can be applied while preserving satisfiability.  For generating
proofs of unsatisfiability, our ability to prove that existential
quantification preserves implication would be sufficient for all of
these cases.  On the other hand, more advanced solvers, such as
SBSAT~\cite{franco-sat-2004}, employ a variety of techniques to prune the
intermediate BDDs based on the structure of other BDDs that remain to
be conjuncted.  This pruning generally reduces the set of satisfying
assignments to the BDD, and so implication does not hold.

In more recent work, we have been able to show that BDD-based methods
can use solution methods that view a Boolean formula as encoding
linear equations over integers or modular
integers~\cite{bryant:tacas:2022}.  Proof-generating BDD operations
can be used to justify the individual steps taken while solving
systems of equations by several different methods.  That has allowed
us to scale the benchmark problems considered in
Section~\ref{sect:experimental} even further, and to avoid the need
for problem-specific solution methods.  We have also demonstrated that
proof-generating BDDs can be integrated into a convention CDCL solver
to allow it to use Gauss-Jordan elimination on the parity constraints
encoded in the formula~\cite{bryant:pos22:cms:2022}.  Overall, we believe
that BDD-based methods can augment other SAT solving methods to
provide new capabilities.

The ability to generate correctness proofs in a BDD-based SAT solver
invites us to also consider generating proofs for other tasks to which BDDs
are applied.  We have already done so for quantified Boolean formulas,
demonstrating the ability to generate proofs for both true and false
formulas in a unified framework~\cite{bryant:cade:2021}.
Other problems of interest include
model checking and model
counting.  Perhaps a proof of unsatisfiability could provide a useful
building block for constructing correctness proofs for these other tasks.

\begin{acks}

We would like to thank Chu-Min Li and Laurent Simon for sharing their
programs for generating the two classes of Urquhart benchmarks
evaluated in Section~\ref{sect:urquhart}.

\end{acks}

\bibliographystyle{plain}

 \bibliography{references}

\end{document}